%% file: distributed_CIA.tex
\documentclass[10pt,journal,twocolumn,final]{IEEEtran}
\usepackage{multind}
\usepackage[T1]{fontenc}
\usepackage{ae,aecompl}
\usepackage[utf8]{inputenc}
\usepackage[english]{babel}
\usepackage{amsmath}
\usepackage{amsfonts}
\usepackage{amssymb}
\usepackage{amsthm}
\usepackage{times}
\usepackage{verbatim, amsbsy}
\usepackage{graphicx}
\usepackage{epsfig}
\usepackage[normalem]{ulem}
\usepackage{color}
\usepackage{fancyhdr}
\usepackage{lastpage}
\usepackage{listings}
\usepackage{transparent}
\usepackage{bm}
\usepackage{url}
\usepackage{import}
\usepackage{algorithm}
\usepackage{algorithmic}

\input{macros}

\makeatletter
\def\url@CIAstyle{%
  \@ifundefined{selectfont}{\def\UrlFont{\sf}}{\def\UrlFont{\small\ttfamily}}}
\makeatother
\newcommand{\executeiffilenewer}[3]{%
\ifnum\pdfstrcmp{\pdffilemoddate{#1}}%
{\pdffilemoddate{#2}}>0%
{\immediate\write18{#3}}\fi%
}

\DeclareMathOperator{\rank}{\mathrm{rank}}

\DeclareMathOperator{\dimV}{\mathrm{dim}}
\DeclareMathOperator{\spanV}{\mathrm{span}}

\DeclareMathOperator*{\argmax}{\mathrm{arg}\,\mathrm{max}}

\newtheorem{theorem}{Theorem}
\newtheorem{definition}[theorem]{Definition}
\newtheorem{proposition}{Proposition}

\newtheorem{corollary}{Corollary}

\newenvironment{thmproof}{{\bf Proof:}}{\hfill\rule{2mm}{2mm}}

\graphicspath{{figures/}}

\begin{document}

\title{A Distributed Approach to Interference Alignment in OFDM-based Two-tiered Networks}

\author{Marco Maso, \IEEEmembership{Student Member, IEEE,} M\'erouane~Debbah,~\IEEEmembership{Senior~Member,~IEEE,} and Lorenzo Vangelista, \IEEEmembership{Senior Member, IEEE}%
\thanks{Copyright (c) 2013 IEEE. Personal use of this material is permitted. However, permission to use this material for any other purposes must be obtained from the IEEE by sending a request to pubs-permissions@ieee.org.}
\thanks{This work was partially supported by the European Commission in the framework of the FP7 Network of Excellence in Wireless COMmunications NEWCOM\# (Grant agreement no. 318306), and by the ERC Starting Grant 305123 MORE (Advanced Mathematical Tools for Complex Network Engineering).}
\thanks{M. Maso and M. Debbah are with the Alcatel-Lucent Chair on Flexible Radio - SUP\'ELEC, Gif-sur-Yvette, France (e-mail: {marco.maso, merouane.debbah}@supelec.fr).}
\thanks{L. Vangelista is with the Department of Information Engineering, University of Padova, Italy (e-mail: lorenzo.vangelista@unipd.it).}}

\maketitle

\begin{abstract}
In this contribution, we consider a two-tiered network and focus on the coexistence between the two tiers at physical layer. We target our efforts on a long term evolution advanced (LTE-A) orthogonal frequency division multiple access (OFDMA) macro-cell sharing the spectrum with a randomly deployed second tier of small-cells. In such networks, high levels of co-channel interference between the macro and small base stations (MBS/SBS) may largely limit the potential spectral efficiency gains provided by the frequency reuse 1. To address this issue, we propose a novel cognitive interference alignment based scheme to protect the macro-cell from the cross-tier interference, while mitigating the co-tier interference in the second tier. Remarkably, only local channel state information (CSI) and autonomous operations are required in the second tier, resulting in a completely self-organizing approach for the SBSs. The optimal precoder that maximizes the spectral efficiency of the link between each SBS and its served user equipment is found by means of a distributed one-shot strategy. Numerical findings reveal non-negligible spectral efficiency enhancements with respect to traditional time division multiple access approaches at any signal to noise (SNR) regime. Additionally, the proposed technique exhibits significant robustness to channel estimation errors, achieving remarkable results for the imperfect CSI case and yielding consistent performance enhancements to the network. 
\end{abstract}
\vspace{-2mm}
\section{Introduction} \label{sec:intro}

Recently, a new standard for mobile communications, i.e., long term evolution advanced (LTE-A) \cite{rpt:3gpp.36.814}, has been developed to guarantee capacity enhancements over current 3G networks up to three times, and satisfy the ever-growing user data demand \cite{tech:cisco11}. To achieve this goal, a new hierarchical approach to network planning has been proposed, where a tier of \linebreak macro-cell base stations is underlaid with a tier of low-power, small-cell (micro/pico/femto) mobile base stations. This \linebreak two-tiered deployment is an attractive solution to improve the capacity of the current networks, thanks to a better average link quality, more efficient usage of spectrum resources and higher spatial reuse (co-channel deployment) \cite{art:chandrasekhar08}. A proliferation of small-cell base stations (SBSs) and data offloading strategies is to be expected for the next future, presumably yielding a two-tiered approach to network design. 

Deployed by end-users, the SBSs are likely to operate in a plug \& play manner and lack a predefined network infrastructure. It is foreseen that a massive SBSs' deployment would unlikely be possible without a significant simplification of the network management paradigms \cite{tech:ngmn08, art:lopezperez2009of}. Self-organizing network (SON) \cite{rpt:3gpp36.300} technologies are seen as potential key factors for the future evolution of mobile networks. On the other hand, a co-channel deployment of self-organizing macro base stations (MBSs) and SBSs would yield high levels of inter-cell interference (ICI), potentially limiting the expected spectral efficiency enhancements \cite{art:lopezperez2009of}. The impact of the ICI on the performance of a general macro-cell based network has been widely studied in the literature \cite{art:gesbert10}. Nevertheless, the nature of the ICI in self-organizing two-tiered networks is twofold. In particular, each standalone base station operating in these networks may generate  \textit{co-tier interference} towards receivers belonging to the same tier, and \textit{cross-tier interference} towards receivers belonging to a different tier. During the standardization phase of recent systems, e.g., LTE-A, \linebreak ICI coordination techniques have been extensively discussed, and are still considered an open problem in the self-configuring and self-optimizing network use cases \cite{rpt:3gpp36.902}. Consequently, the new network paradigms require not only the design of new protocols to allow simplified network operations, but also the study of novel signal processing techniques to provide the expected spectral efficiency gains at physical layer \cite{onl:artist4g10, art:delima12}. 

Several state-of-the-art coordinated interference management techniques have been proposed in the literature, to realize the coexistence of the two tiers and enhance the spectral efficiency of such networks. Interference alignment (IA) based solutions \cite{art:cadambe2008i} conceal the interference at the unintended receiver in a signal subspace of constant size, and can be implemented for several channel state information (CSI) assumptions. This is achievable if exploitable degrees of freedom are available in time, frequency or spatial domain, and always requires a peculiar decoding at the receiver to realize the alignment. Conversely, coordinated beamforming approaches \cite{art:gesbert10} involve signal processing at the transmitter only, but have more stringent CSI and signaling constraints. On the other hand, if the two tiers do not cooperate, the second tier is in general considered as subordinated to the first tier and opportunistic transmit strategies are likely to be adopted by the SBSs \cite{art:delima12}. Proposed to promote an opportunistic usage of the available resources, cognitive radio (CR) \cite{art:haykin2005} approaches can be adopted to frame a scenario where the SBSs may sense the environment to change and adapt their transmit parameters accordingly \cite{art:akoum11, art:zhang10a}. By denoting the first tier as the primary system and the second opportunistic tier as the secondary system, any two-tiered network can be easily framed according to the CR paradigm \cite{art:lopezperez2009of}. Several solutions based on IA or transmit beamforming have been proposed for the CR setting \cite{art:perlaza2009, art:gastpar07, art:shen11}, usually requiring several degrees of cross-tier and co-tier coordination and multiple spatial dimensions at the transmitter and/or receiver. It is important to note that all the aforementioned techniques involve a bi-directional signaling between the MBSs/SBSs to be implemented, requiring the existence of a link to this scope, e.g., X2 interfaces \cite{rpt:3gpp36.300} and/or dedicated backhauls. However, due to the massive and unplanned SBSs' deployment, this link may neither exist nor provide sufficient performance to meet the latency and delay requirements of any of the discussed state-of-the-art techniques \cite{onl:artist4g11}. 

In this contribution, we consider a two-tiered network composed of two independent tiers, with frequency \linebreak reuse 1 and operating in time division duplex (TDD) mode. We focus on a LTE-A orthogonal frequency division multiple access (OFDMA) macro-cell (first tier) with a single antenna MBS serving a group of macro user equipments (MUEs), coexisting with several cognitive self-organizing SBSs (second tier) serving one small user equipment (SUE) each. We start from the previously drawn insights and propose a completely self-organizing approach to cross- and co-tier interference management at the SBSs that does not require the X2 interface to operate, i.e., no explicit cooperation and signaling in the second tier or among tiers is established. Accordingly, in contrast to the aforementioned works, we do not assume that the SBSs have access to information about the \linebreak MBS' transmitted message, power allocation, or about the existence of available space, time or frequency degrees of freedom in the first tier. A novel cognitive interference alignment (CIA) scheme is introduced to realize the coexistence between the two tiers and yield an overall spectral efficiency enhancement for the network. Numerical analysis show that a local input signal subspace reduction, relying only on the information about the number of neighboring SUEs, is sufficient to each SBS to derive the desired precoder autonomously. Remarkably, the obtained results demonstrate that only CSI w.r.t. the link towards the served SUE is necessary at the SBSs to cope adequately with the co-tier interference, and CSI w.r.t. the links towards the non-served SUEs does not provide spectral efficiency improvements to the second tier. CIA is shown to outperform user orthogonalization approaches not requiring special signaling between the SBSs, e.g., time/frequency division multiple access (TDMA/FDMA), in the sense of achievable spectral efficiency. Additionally, the impact of imperfect CSI at the SBSs is studied, and the proposed technique exhibits significant robustness to channel estimation errors, yielding consistent performance enhancements to the two-tiered network at any signal to noise ratio (SNR).

This paper is organized as follows. In Sec. \ref{sec:model}, the scenario and signal model is introduced. The cognitive interference alignment scheme is described in Sec. \ref{sec:cia}. The optimal linear transmit strategy is derived in Sec. \ref{sec:precoder_c}. A discussion on the spectral efficiency of the considered network is proposed in Sec. \ref{sec:spec}. The performance of the proposed scheme is evaluated numerically in  Sec. \ref{sec:num_anal}. Finally, in Sec. \ref{sec:conclusion}, conclusions and future research directions are presented.

\section{Model} \label{sec:model}

\begin{figure}[!h]
	\centering
	\includegraphics[width=.99\columnwidth]{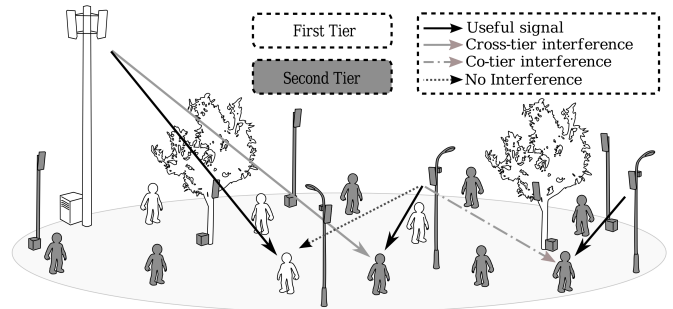}
	\caption{Two-tiered self-organizing network [downlink].}	
	\label{fig:scenario}
\end{figure}

Consider the downlink of a two-tiered network as depicted in Fig.~\ref{fig:scenario}, where, for clarity, only a reduced number of channels are represented.
In compliance with the supported transmit modes in 4G standard as LTE/LTE-A \cite{rpt:3gpp.36.814, rpt:3gpp25.814}, we assume that the communications in the two tiers are performed in TDD mode. In the first tier, a group of $M$ MUEs is served by a licensee single antenna MBS, by means of an OFDMA transmission \cite{rpt:3gpp25.814}. A second tier, comprised of $K$ single antenna cognitive SBSs, is deployed inside the same area. Frequency reuse 1 is adopted for matters of spectral efficiency, thus each SBS opportunistically transmits over the same bandwidth as the MBS. For clarity and simplicity in the model we assume that each SBS serves only one SUE. However, this does not decrease the generality of the approach. An extension to a multi-SUEs per SBS model could be seamlessly obtained by means of any multi-user scheduling technique \cite{art:andrews01}, once the solution for single SUE case has been identified. The first legacy tier is oblivious of the existence of the second, thus the two tiers are completely independent and no cross-tier cooperation is established. Therefore, the MBS does not implement any interference mitigation strategy, whereas according to the CR paradigm, the SBSs must protect the MUEs from undesired cross-tier interference. In the considered scenario, every receiver in the system is potentially able to act as a MUE or SUE, depending on its network association, thus both act as a classic OFDM receiver. On the other hand, concerning the transmitters, we assume that both systems adopt Gaussian constellations.

Concerning the notation, in this work we let $\Id_M$ be the \linebreak $M \times M$ identity matrix and $\zerov_{N,M}$ be the $N \times M$ all zeros matrix. We define $[\Am]_{m,n}$ as the element of the matrix $\Am$ at the $m$th row and the $n$th column. Given a vector \linebreak $\av=(a_1,\dots,a_N)$, we denote as $d(\av)=diag(\av)$ a diagonal matrix such that $[d(\av)]_{i,i}=a_i$, for the sake of compactness of the notation. The expectation operator is defined as $\mathbb{E}$, whereas $\ker(\Am)$ denotes the kernel of the matrix $\Am$ and $\trace(\Am)$ its trace. All vectors are columns, unless otherwise stated. Throughout the paper, in vector and matrix definitions, subscript ``$\text{p}$'' and ``$\text{s}$'' refers to the primary and secondary system respectively. Moreover, the superscript $(i,j)$ denotes the pair composed of transmitter $i$ and receiver $j$, such that $h_{\text{sp}}^{(i,j)}$ represents the channel from $i$th SBS to the $j$th MUE. Finally, given two sets $\Ac$, $\Bc$, such that $\Ac \subset \Bc$, we define $1{\hskip -2.5 pt}\hbox{I}_{\Ac}\Bc$ as the indicator function of the subset $\Ac$ of the set $\Bc$. 

Let $\hv_{\text{a}\text{b}}^{(\cdot,\cdot)}\sim{\Cc}\Nc(0,\Id_{l+1}/(l+1))$ be i.i.d. Rayleigh fading channel vectors of size $l+1$ taps. We assume for simplicity that the MBS uniformly assigns the $N$ active subcarriers to the MUEs and that the cyclic prefix, added by the MBS to the OFDM block to combat inter-symbol (ISI) and inter-block interference (IBI), is of size $L$, such that $L > l$. Therefore, the primary system transmitted block size is $N+L$. Let $\sv_{\text{p}}\sim\Cc\Nc(0, d(p_{\px,1},\dots,p_{\px,N}))$ be the $N$-sized MBS' input symbol vector. Then, if we define $\Fm \in \mathbb{C}^{N\times N}$ as a unitary discrete Fourier transform (DFT) matrix with \linebreak $[\Fm]_{(k+1, l+1)}=\frac{1}{\sqrt{N}}e^{-i2\pi \frac{kl}{N}}$ for $k,l = \{ 0,\dots,N-1\}$ and $\Am$ a $(N+L)\times N$ cyclic prefix insertion matrix given by
\begin{equation*}
 \Am = \left[
\begin{array}{c}
 \mathbf{0_{L,N-L}}\quad \Id_L \\
\Id_{N}
\end{array}
\right],
\end{equation*}
we can write the MBS' transmit symbol vector, of size $N+L$, as
\begin{equation}
\xv_{\px} = \Am \Fm^{-1} \sv_{\text{p}}.
\end{equation}
For consistency, we assume that each SBS adopts a block transmission scheme as the MBS, i.e., block size of $N+L$, hence we define $\xv_{\sx}^{(i)} \in \mathbb{C}^{(N+L) \times 1}$ as the $i$th SBS' transmit symbol vector. Now let $\hv_{\text{ab}}^{(\cdot,\cdot)}=(h_{\ax\bx,0}^{(\cdot,\cdot)},\dots,h_{\ax\bx,l}^{(\cdot,\cdot)})$ be a channel vector representing the link between a transmitter in the tier "a" and a receiver in the tier "b", thus we can express its channel circulant matrix $\Hm_{\ax \bx}^{(\cdot,\cdot)} \in \mathbb{C}^{(N+L)\times (N+L)}$ as
\begin{equation*}
\Hm_{\ax\bx}^{(\cdot,\cdot)} =
\left[ \begin{array}{ccccccc}
h_{\ax\bx,0}^{(\cdot,\cdot)} & 0 & \cdots & h_{\ax\bx,l}^{(\cdot,\cdot)} & \cdots & \cdots & h_{\ax\bx,1}^{(\cdot,\cdot)} \\
\vdots & \ddots & \ddots &  & \ddots & \ddots & \vdots \\
\vdots & \ddots & \ddots &  & \ddots & \ddots & h_{\ax\bx,(l-1)}^{(\cdot,\cdot)} \\  
h_{\ax\bx,l}^{(\cdot,\cdot)} & \cdots & \cdots & h_{\ax\bx,0}^{(\cdot,\cdot)} & 0 & \cdots & 0\\
0 & \ddots & \ddots & \ddots & \ddots & \ddots & \vdots \\
\vdots & \ddots & \ddots & \ddots & \ddots & \ddots & 0 \\ 
0 & \ddots & \ddots & h_{\ax\bx,l}^{(\cdot,\cdot)} & \ddots & \ddots & h_{\ax\bx,0}^{(\cdot,\cdot)}\\ 
 \end{array} \right],
\end{equation*}
where, as an approximation, we neglect the inter-channel interference that may be caused by analog and radio frequency impairments at the MUEs/SUEs. Let us consider the $j$th MUE. We define  $\Bm \in \mathbb{C}^{N \times (N+L)}$ as a cyclic prefix removal matrix given by
\begin{equation*}
 \Bm = \left[
\begin{array}{c}
 \mathbf{0_{N,L}}\quad \Id_N \\
\end{array}
\right],
\end{equation*}
hence we can write $\yv_{\px}^{(j)} \in \mathbb{C}^{N \times 1}$, received signal at $j$th MUE as
\begin{equation}\label{eq:received_p}
    \yv_{\px}^{(j)} = \Fm \Bm  \left(\Hm_{\text{pp}}^{(1,j)}\Am \Fm^{-1}\xv_{\px} + \sum_{i=1}^{K} \Hm_{\text{sp}}^{(i,j)}\xv_{\sx}^{(i)} + \nv_{\px}^{(j)}\right),
\end{equation}
where $\nv_{\px}^{(j)}\sim\Cc\Nc(0, \sigma^2 \Id_{N+L})$ is an additive white Gaussian noise (AWGN) vector and $\sum_{i=1}^{K} \Hm_{\text{sp}}^{(i,j)}\xv_{\sx}^{(i)}$ is the cross-tier interference generated by the SBSs. Analogously, we can write $\yv_{\sx}^{(k)} \in \mathbb{C}^{N \times 1}$, received signal at $k$th SUE as
\begin{IEEEeqnarray}{rCl}
    \yv_\sx^{(k)} &=& \Fm \Bm \bigg(\Hm_{\text{ps}}^{(1,k)}\Am \Fm^{-1}\xv_{\px} + \Hm_{\text{ss}}^{(k,k)}\xv_{\sx}^{(k)} + ... \nonumber \\
    && +\: \sum_{i=1, i \neq k}^{K} \Hm_{\text{ss}}^{(i,k)}\xv_{\sx}^{(i)} + \nv_{\sx}^{(k)}\bigg), \label{eq:received_s}
\end{IEEEeqnarray}
%
where $\nv_{\sx}^{(k)}\sim\Cc\Nc(0, \sigma^2 \Id_{N+L})$ is an AWGN vector. Note that, in \eqref{eq:received_s}, $\Hm_{\text{ss}}^{(k,k)}\xv_{\sx}^{(k)}$ is the received signal coming from the $k$th SBS and $\Hm_{\text{ss}}^{(i,k)}\xv_{\sx}^{(i)}$ the co-tier interference component coming from the other SBSs.

\section{Cognitive interference alignment} \label{sec:cia}

An opportunistic secondary tier can perform its operations according to different policies, depending on the adopted spectrum access approach. Accordingly, its cross-tier interference towards the licensee tier can be canceled, i.e., \textit{overlay} CR network \cite{art:devroye2006}, or mitigated,  i.e., \textit{underlay} CR network \cite{art:ma2010}. On the other hand, if both tiers act opportunistically, adaptively accessing the spectrum to avoid collisions, an \textit{interweave} CR network is realized \cite{art:haykin2005}. The assumptions made in Sec. \ref{sec:model} clearly imply that the considered scenario is modeled according to the overlay paradigm, hence a technique to null the cross-tier interference coming from the SBSs has to be devised. We assume that no spectrum sensing is performed in the second tier, and no information about the spectrum characteristic, time resource allocation, primary system's message and power allocation is available at the SBSs. Thus, techniques based on dirty paper coding (DPC) \cite{art:costa83} or opportunistic IA \cite{art:perlaza2009} cannot be implemented. Furthermore, due to the absence of a backhaul in the second tier, and the lack of information about the presence or performance of the X2 interface, no signaling is exchanged among SBSs. Consequently, techniques relying on coordinated beamforming at the transmitter are not implementable, and each SBS must adopt self-configuring and self-optimizing procedures. Finally, no cooperation is established between the SUEs/MUEs, i.e., single user decoding is performed. This is made to frame a scenario that does not rely on too unrealistic assumptions, as well as on hardly practically implementable algorithms in terms of required time and computational capabilities.

On the other hand, the TDD mode assumed in Sec. \ref{sec:model} provides a structured transmission scheme that can be opportunistically exploited in the second tier to enhance the network performance. This mode has raised an increasing interest in the research community as the key factor for many \linebreak state-of-the-art technological advancements, e.g., massive (or network) multiple input multiple output (MIMO) \cite{art:huh12}, to provide significant spectral efficiency gains w.r.t. legacy frequency division duplex (FDD) mode approaches (the interested reader may refer to \cite{art:chan06} and references therein for further information on the TDD/FDD topic). In our scenario, each opportunistic SBS can capitalize on the structure of the TDD communication to design a suitable interference management strategy. In particular, information such as local CSI w.r.t. the links towards the SUEs/MUEs reached by its transmission, and knowledge about the downlink physical resource blocks (PRBs) \cite{rpt:3gpp25.814} allocation performed by the MBS in the first tier, may be acquired. One possible way to perform the channel estimation in LTE/LTE-A scenarios has been proposed in \cite{conf:bert11}, where the TDD structure of the transmission and the presence of the sounding reference signals in the LTE/LTE-A frame \cite{rpt:3gpp25.814} are exploited to estimate the channel, considered reciprocal throughout the duration of the coherence time. Concerning the PRBs allocation in LTE-A, this is communicated by the MBS to the MUEs through the physical downlink control channel (PDCCH), within each downlink sub-frame \cite{rpt:3gpp36.211}. This information is meant to reach all the MUEs hosted in the cell, thus it is received by the SBSs as well, being the latter deployed in the coverage area of the MBS. 

Concerning the model introduced in Sec. \ref{sec:model}, we would like to note that, this contribution is the first step towards a comprehensive characterization of the performance of a \linebreak two-tiered network under perfect and imperfect CSI assumption. Accordingly, herein we put focus on the achievable ultimate bounds in case of a first tier composed by a single MBS, whereas no limit is imposed on the number of deployed SBSs. The rationale for this is that if the proposed solution could not provide meaningful performance even in a single MBS case (for perfect and imperfect CSI), then there would be no use in pursuing the analysis for more complex cellular layouts including multiple MBSs, structured SBSs' positioning and practical channel models. 

In the following, we will see how a suitable linear transmit scheme for the SBSs based on IA~\cite{art:cadambe2008i} naturally arises from the network configuration and constraints. In Sec. \ref{sec:cross_tier} the cognitive interference alignment (CIA) scheme is introduced as an inner precoder to null the cross-tier interference towards the legacy system. Afterwards, the design of an outer precoder to manage the co-tier interference, yielding promising spectral efficiency values for the second link, is presented in Sec. \ref{sec:co_tier}. 

\subsection{Cross-tier interference alignment} \label{sec:cross_tier}

In an OFDMA system, each receiver performs a simple base-band processing after the DFT operation performed in \eqref{eq:received_p}. Let $\Nc$ be the set of active subcarriers, and $\Nc_{j}$, be the set of $\frac{N}{M}$ subcarrier indices assigned to the $j$th MUE, i.e., one or more PRBs, with $\displaystyle \bigcup_{j=1}^{M}\Nc_{j} = \Nc$ and $\displaystyle \bigcap_{j=1}^{M}\Nc_{j} = \emptyset$ by construction. Moreover, let $\Dm_j \in \{0,1\}^{(N \times N)}$ be the filter adopted by the $j$th MUE to recover $\Nc_j$, with $[\Dm_j]_{(n, n)} = 1$ if $n \in \Nc_{j}$ and zero otherwise, such that
\begin{equation} \label{eq:Dmj}
\begin{cases}
\displaystyle\sum_{n=1}^{N} 1{\hskip -2.5 pt}\hbox{I}_{\{n \in \Nc_{j}\}}[\Dm_j]_{(n, n)}~=\frac{N}{M} \\
\displaystyle\sum\limits_{j=1}^M \Dm_j = \Id_N.
\end{cases}
\end{equation}
Therefore, the $j$th MUE can extract the PRBs of interest by means of $\Dm_j$ and we can rewrite \eqref{eq:received_p} as 
\begin{equation} \label{eq:y_p_Dmj}
\yv_{\px}^{(j)} = \Dm_j \Fm  \Bm \left(\Hm_{\text{pp}}^{(1,j)}\Am \Fm^{-1}\xv_{\px} + \nv_{\px}^{(j)}\right) + \sum_{i=1}^{K} \Tm_{\sx\px}^{(i,j)} \xv_{\sx}^{(i)},
\end{equation}
with $\Tm_{\sx\px}^{(i,j)} = \Dm_j \Fm \Bm  \Hm_{\text{sp}}^{(i,j)} \in \mathbb{C}^{N \times (N+L)}$ equivalent representation of the interference link from the $i$th SBS to the $j$th MUE. In particular, this operation realizes the user orthogonality in the frequency domain, thus an equivalent representation of the overall received signal in the first tier can be found from \eqref{eq:y_p_Dmj} as
\begin{align}
\begin{split}
\yv_\px &= \sum_{j=1}^{M} \yv_{\px}^{(j)} = \sum_{j=1}^{M}\Dm_j \Fm \Bm   \bigg(\Hm_{\text{pp}}^{(1,j)}\Am \Fm^{-1}\xv_{\px} + ... \\
& \hspace{20mm} + \sum_{i=1}^{K} \Hm_{\text{sp}}^{(i,j)}\xv_{\sx}^{(i)} + \nv_{\px}^{(j)}\bigg).
\end{split} \label{eq:overall_yp}
\end{align}
As a consequence, we can define $\Tm_{\sx\px}^{(i,\cdot)} = \sum_{j=1}^{M} \Tm_{\sx\px}^{(i,j)}$ as the overall interference channel between the $i$th SBS and the MUEs in the first tier. In the considered scenario, by construction, $\rank({\Tm_{\sx\px}^{(i,\cdot)}}) = N$, thus for the rank-nullity theorem we have
\begin{equation} \label{eq:rank_null}
\dimV \ker(\Tm_{\sx\px}^{(i,\cdot)})= L,
\end{equation}
$\forall \Hm_{\text{sp}}^{(i,j)} \in \mathbb{C}^{(N+L) \times (N+L)}$. If we assume perfect CSI at the SBSs w.r.t. the cross-tier interference link $\hv_{\sx\px}^{(\cdot,\cdot)}$, the $i$th SBS can always find a matrix $\Em^{(i)} \in \mathbb{C}^{(N+L) \times L}$ such that 
\begin{equation} \label{eq:nullcondition}
    \spanV (\Em^{(i)}) = \ker(\Tm_{\sx\px}^{(i,\cdot)})\footnote{Let $\Vc_1$ and $\Vc_2$ be two vector spaces of dimension $M$. We define $\Vc_1 = \Vc_2$ if and only if $\forall x \in \mathbb{C}^{M}$, $x \in \Vc_1 \leftrightarrow x \in \Vc_2$.},
\end{equation}
and $\Tm_{\sx\px}^{(i,\cdot)} \Em^{(i)} =\textbf{0}_{N,L}$. In an IA based transmission, the overall received signal space has to be decomposable in two constant sized components, an interference and a useful signal subspace. Now, let us introduce $\sv_{\sx}^{(i)}\sim\Cc\Nc(0, d(p^{(i)}_{\sx,1},\dots,p^{(i)}_{\sx,L}))$ as an $L$-sized input symbol vector at the $i$th SBS, and define
\begin{equation} \label{eq:x_E}
\xv_\sx^{(i)} = \Em^{(i)} \sv_\sx^{(i)}.
\end{equation}
For the sake of clarity, we keep our focus on the $j$th MUE and let $\tilde{\Tm}_{\sx\px}^{(i,j)} = \Tm_{\sx\px}^{(i,j)} \Em^{(i)} \in \mathbb{C}^{N \times L}$. We can rewrite \eqref{eq:y_p_Dmj} as 
\begin{equation} \label{eq:y_p_bef_proc}
\yv_{\px}^{(j)} = \Dm_j \Fm  \Bm \left(\Hm_{\text{pp}}^{(1,j)}\Am \Fm^{-1}\xv_{\px} + \nv_{\px}^{(j)}\right) + \sum_{i=1}^{K} \tilde{\Tm}_{\sx\px}^{(i,j)} \sv_{\sx}^{(i)},
\end{equation}
where $\tilde{\Tm}_{\sx\px}^{(i,j)}$ has the $\frac{N}{M}$ rows whose indexes $n \in \Nc_j$ composed of zero entries. Consequently, we can write
\begin{equation} \label{eq:eq_rx_ch_p}
\sum_{n \in \Nc_j} \sum_{m=1}^{L}[\tilde{\Tm}_{\sx\px}^{(i,j)}]_{(n,m)} = 0,
\end{equation}
and this holds $\forall \sv_\sx^{(i)} \in ~\mathbb{C}^{L \times 1}$, regardless of the size of $\hv_{\sx\px}^{(i,j)}$, according to \eqref{eq:Dmj} and \eqref{eq:nullcondition}. The interference signal coming from $i$th SBS is aligned at the $j$th MUE, and confined into the same constant sized subset of subcarriers given by $\Nc \setminus \Nc_j$, $\forall i \in [1,K]$. As a consequence, the desired $\frac{N}{M}$ interference free dimensions at the $j$th MUE can be obliviously extracted by processing the received signal in \eqref{eq:received_p} with $\Dm_j$, as in the classic OFDMA receiver processing, to finally obtain
\begin{equation} \label{eq:y_p_aft_proc}
\yv_{\px}^{(j)} = \Dm_j \Bm \Fm \left(\Hm_{\text{pp}}^{(1,j)}\Am \Fm^{-1}\xv_{\px} + \nv_{\px}^{(j)}\right),
\end{equation}
where the cross-tier interference coming from the second tier has been completely canceled. At this stage, we can define the spectral efficiency for the $j$th MUE as
\begin{equation} \label{eq:spec_eff_Rp}
R_{\px}^{(j)} = \frac{1}{N+L}\sum_{i=1}^N \log_2 \left(1 + \text{SINR}_{(\px,i)}^{(j)}\right),
\end{equation}
with
\begin{equation} \label{eq:SINR_p}
\text{SINR}_{(\px, i)}^{(j)} = \frac{p_{\px,i} \left|\left[\Dm_j \Fm  \Bm \Hm_{\text{pp}}^{(1,j)}\Am \Fm^{-1}\right]_{(i,i)}\right|^2}{\sigma_n^2},
\end{equation} 
signal to interference plus noise ratio (SINR) of its $i$th received symbol. We remark that the choice of $\Dm_j$ depends uniquely on the resource allocation performed at the MBS. As a consequence, the degrees of freedom of the primary OFDMA transmission and its overall spectral efficiency $R_{\px}=\displaystyle \sum_{j=1}^M R_{\px}^{(j)}$ are preserved, regardless of the transmit power at the $i$th SBS. 

Switching the focus on the second tier, we have shown that the legacy system's characteristics, such as the OFDMA receiver processing and the redundancy introduced by the MBS to combat ISI and IBI, provide a constant number of transmit opportunities for the SBSs. Each of the $K$ SBSs can potentially exploit up to $L$ degrees of freedom, $\forall K \in \mathbb{N}$, thus the maximum size of $\sv_\sx^{(i)}$ is $1 \times L$. Nevertheless, the multi-user interference generated by concurrent transmissions in the second tier may strongly limit the achievable spectral efficiency of the secondary system. In the following section a strategy to deal with this co-tier interference is introduced. The overall performance of the second tier will be discussed in Sec. \ref{sec:num_anal}.

\subsection{Co-tier interference mitigation} \label{sec:co_tier}

The CIA scheme presented in Sec. \ref{sec:cross_tier} guarantees up to $L$ transmit dimensions per SBS. The interference alignment and consequent nulling can be obtained regardless of $K$, number of SBSs. This remarkable achievement shows the potential of CIA when perfect CSI w.r.t. the cross-tier interference link is available in the second tier. Nevertheless, this imposes hard constraints for a practical implementation of this scheme. In realistic scenarios, a channel estimation is valid only throughout the duration of the coherence time of the channel, which is finite. As a consequence, fast and preferably one-shot strategies to mitigate the co-tier interference are needed, and approaches relying on infinite channel extensions are not implementable \cite{conf:zhou11d}. Additionally, since the SUEs are standard OFDM receivers, no further decoding for subspace decomposition to realize the IA can be adopted. Thus, a fundamental requirement to implement standard distributed IA schemes relying on iterative schemes between transmitter and receiver \cite{conf:zhou10d} is not met. The distributed approach relying only on transmitter processing for downlink multi-cell scenarios proposed in \cite{art:ho11} cannot be adopted as well. Despite the promising achievable spectral efficiency, this approach is hardly applicable in our scenario, since it is DPC-based, thus affected by practical implementability issues. As assumed in Sec. \ref{sec:model}, each SBS disposes of only one antenna, hence, due to dimensionality issues, i.e., $N >> L$, no coordinated \cite{art:gesbert10} or distributed beamforming \cite{art:ho11} can be performed efficiently. Finally, as discussed in Sec. \ref{sec:intro}, due to the unplanned  SBSs' deployment, no bi-directional signaling can be guaranteed in the second tier, and the co-tier interference problem must be addressed by means of a self-organizing technique implementable by the SBSs in an autonomous manner. Accordingly, in this section we seek for a linear outer precoder to be adopted by the SBSs to provide the desired spectral efficiency enhancements for the considered two-tiered network, and a novel distributed interference management scheme is proposed. 

We note that, the co-tier interference in the second tier is completely unrelated to the cross-tier interference generated by the MBS towards the SUEs. Regardless of the number of active SBSs, the latter will always impact the performance of the second tier and does not represent a parameter that can be appropriately tuned by means of a mechanism at the SBSs. Consequently, in this section the equations will not include the cross-tier interference coming from the first tier. We remark that this simplification is made to simplify the equations in the algorithm derivation and does not imply the assumption of absence of cross-tier interference from the MBS to SUEs. We start by plugging the previously derived CIA precoder into \eqref{eq:received_s}. We focus on the $k$th SUE and write
\begin{IEEEeqnarray}{rCl}
 \yv_\sx^{(k)} &=& \Fm \Bm \bigg(\Hm_{\text{ss}}^{(k,k)}\Em^{(k)} \sv_{\sx}^{(k)} + ... \nonumber \\ && +\: \sum_{i=1, i \neq k}^{K} \Hm_{\text{ss}}^{(i,k)} \Em^{(i)} \sv_{\sx}^{(i)} + \nv_{\sx}^{(k)}\bigg).
 \label{eq:received_s2}
\end{IEEEeqnarray}
Now, let
\begin{equation}
\Tm_{\sx\sx}^{(j,k)} = \Fm \Bm \Hm_{\text{ss}}^{(j,k)}\Em^{(j)} = [\tv_{\sx\sx,1}^{(j,k)} \hspace{1mm} | \hspace{1mm} \dots \hspace{1mm} | \hspace{1.8mm} \tv_{\sx\sx,N}^{(j,k)}]^{\text{T}} \in \mathbb{C}^{N \times L}
\end{equation}
be the equivalent representation of the channel between the $j$th SBS and the $k$th SUE, $\forall j \in [1,K]$. Note that, $\Fm$ is unitary hence $\tilde{\nv}_\sx = \Fm \Bm \nv_{\sx}^{(k)}\sim\Cc\Nc(0, \sigma^{2}\Id_{N})$. After the DFT, the $k$th SUE disposes of $N$ received symbols, i.e., $N$ different linear combinations of the $L << N$ input symbols transmitted by the $k$th SBS, corrupted by the interference coming from the remaining $K-1$ SBSs and by the thermal noise. As a consequence, the equivalent CIA channel representation provides a significant receive diversity effect to the second tier transmission. Interestingly, the SBSs may exploit this inherent feature of the system to induce a power gain at each SUE \cite{book:tse2005f}, with no need for cooperation or coordinated beamforming strategies. Consider the $i$th received symbol at $k$th SUE, its SINR is defined as
\begin{equation} \label{eq:SINR_ki}
\text{SINR}^{(k)}_{(\sx,i)} = \frac{\tv_{\sx\sx,i}^{(k,k)\text{H}}d(p^{(k)}_{\sx,1},\dots,p^{(k)}_{\sx,L})\tv_{\sx\sx,i}^{(k,k)}}{\sigma^2 + \sum_{j=1, j \neq k}^{K} \tv_{\sx\sx,i}^{(j,k)\text{H}}d(p^{(j)}_{\sx,1},\dots,p^{(j)}_{\sx,L})\tv_{\sx\sx,i}^{(j,k)}}.
\end{equation}
Moreover, we let $R_{\sx}=\displaystyle \sum_{k=1}^K R_{\sx}^{(k)}$ be the spectral efficiency of the second tier, where
\begin{equation} \label{eq:spec_eff_R}
R_{\sx}^{(k)} = \frac{1}{N+L}\sum_{i=1}^N \log_2 \left(1 + \text{SINR}^{(k)}_{(\sx,i)}\right)
\end{equation}
is the spectral efficiency of the link between the $k$th SBS/SUE pair, logarithmic function of $\text{SINR}^{(k)}_{(\sx,i)}$. Diversity and power gains directly impact the SINR at the receiver. Therefore, the input signal subspace dimension at the SBSs, affecting the dimensionality of the system, has a fundamental role in the performance of the second tier. 

No communication is established between the SBSs, hence no cooperation can be realized. However, an implicit coordination mechanism could be adopted to aim at an overall network spectral efficiency enhancement. In particular, we can design a transmit strategy that constrains the input symbol subspace to belong to $\mathbb{C}^{1 \times \theta}$, with $\theta \in [1,L]$ natural number. Accordingly, the dimensionality of the equivalent channel seen by the $k$th SBS may adaptively change depending on the number of neighboring SUE/SBS pairs, to improve the receive diversity, thus the SINR per received symbol. Let $\Thetam^{(k)} \in \mathbb{C}^{L \times \theta}$ be an outer precoder such that $\sv_\sx^{(k)}=\Thetam^{(k)}\uv_\sx^{(k)}$ and $\Zm^{(k)} = \Em^{(k)}\Thetam^{(k)} \in \mathbb{C}^{L \times \theta}$, then \eqref{eq:x_E} can be rewritten as
\begin{equation} \label{eq:x_theta}
\xv_\sx^{(k)} = \Zm^{(k)} \uv_\sx^{(k)},
\end{equation}
where $\uv_\sx^{(k)}\sim\Cc\Nc(0, d(p^{(k)}_{\sx,1},\dots,p^{(k)}_{\sx,\theta}))$ is a $\theta$-sized input symbol vector at the $k$th SBS.

At this stage, each SBS has two degrees of freedom to design $\Thetam^{(k)}$, i.e., the choice of a suitable signal subspace structure and its dimension $\theta$. We first focus on the former and assume that $\theta$ is known at each SBS. We remark that, the eigenvalue and eigenvector distribution of a finite dimension Toeplitz matrix is currently not known, and the same holds true for its kernel structure, e.g. $\ker(\Tm_{\sx\px}^{(i,\cdot)})$. Thus, neither deterministic nor stochastic information about the interference generated to the $k$th SUE by the neighboring non-serving SBSs is available. On the other hand, by looking at \eqref{eq:SINR_ki}, we see that the signal transmitted by each SBS contributes to $\text{SINR}^{(k)}_{(\sx,i)}$ either at the numerator or at the denominator. Consequently, only two possible strategies can be identified to increase $R_{\sx}^{(k)}$, i.e., the $k$th SBS contribution to $R_{\sx}$, as follows.

\subsubsection{CIA A}

The first approach, hereafter denoted as \textit{CIA A}, is an aggressive strategy adopted by the $k$th SBS aiming at the maximization of \eqref{eq:spec_eff_R}, without considering the impact on the denominator of \eqref{eq:SINR_ki} experienced by the SINR at the non-served SUEs. The scenario reduces to a \linebreak point-to-point link connecting the $k$th SBS/SUE pair disturbed by both the interference generated by the non-serving SBSs and AWGN. The interference generated by the non-serving SBSs to the $k$th SUE cannot be predicted by the $k$th SBS, thus it is ignored and only the CSI related to $\hv_{\text{ss}}^{(k,k)}$ is needed to perform the input subspace selection. Let $\Tm_{\sx\sx}^{(k,k)}=\Um_{\Tm}^{(k)} \Lambdam_{\Tm}^{(k)} \Vm_{\Tm}^{(k)}$ be the singular value decomposition (SVD) of the equivalent channel connecting the $k$th SBS/SUE pair, with $\Um_{\Tm}^{(k)}=[\uv_{\Tm,1}^{(k)} | \dots | \uv_{\Tm,N}^{(k)}] \in \mathbb{C}^{N \times N}$, $\Vm_{\Tm}^{(k)}=[\vv_{\Tm,1}^{(k)} | \dots | \vv_{\Tm,L}^{(k)}] \in \mathbb{C}^{L \times L}$ unitary matrices, and with
\begin{equation}
\Lambdam_{\Tm}^{(k)}=\left[d\left(\sqrt{\lambda_{\Tm,1}^{(k)}},\dots,\sqrt{\lambda_{\Tm,L}^{(k)}}\right), \textbf{0}_{L,N-L}\right]^{T},
\end{equation}
where $\sqrt{\lambda_{\Tm,i}^{(k)}}$ are the non-negative singular values of $\Tm_{\sx\sx}^{(k,k)}$ such that $\sqrt{\lambda_{\Tm,1}^{(k)}} \geq \sqrt{\lambda_{\Tm,2}^{(k)}} \geq \dots \geq \sqrt{\lambda_{\Tm,L}^{(k)}}$. Then the $k$th SBS can approximate \eqref{eq:spec_eff_R} as a sum of decreasing positive terms and write
\begin{equation} \label{eq:spec_eff_R2}
\widehat{R}_{\sx}^{(k)} = \frac{1}{N+L}\sum_{i=1}^{L} \log_2{(1 + \frac{\lambda_{\Tm,i}^{(k)}}{\sigma^2})}.
\end{equation}
A this stage, we can define $\Theta^{(k)} = [\vv_{\Tm,1}^{(k)} | \dots | \vv_{\Tm,\theta}^{(k)}] \in \mathbb{C}^{L \times \theta}$, with $\theta \in [1,L]$, as the outer precoder that aligns the input signal to the $\theta$ strongest eigenmodes of $\Tm_{\sx\sx}^{(k,k)}$, i.e., the most efficient subspace of $\spanV{(\Em^{(k)})}$ for the direct link spectral efficiency maximization. Then, \eqref{eq:spec_eff_R2} can be rewritten as
\begin{equation} \label{eq:spec_eff_R2_new}
\widehat{R}_{\sx}^{(k)}(\theta) = \frac{1}{N+L}\sum_{i=1}^{\theta} \log_2{(1 + \frac{\lambda_{\Tm,i}^{(k)}}{\sigma^2})},
\end{equation}
where the dependency of the spectral efficiency on $\theta$ has been specified for clarity. Note that, in  \eqref{eq:spec_eff_R2} and \eqref{eq:spec_eff_R2_new}, a uniform power allocation at the $k$th SBS has been assumed for simplicity in the representation, with $p_{\sx,i}^{(k)} = 1$, $\forall i \in [1,L]$. This does not reduce the generality of the approach. In fact, for a given power budget, stronger eigenmodes always result in a more efficient transmission, regardless of the power loading strategy \cite{book:tse2005f}. 

\subsubsection{CIA B}

The second approach, hereafter denoted as \textit{CIA B}, is a conservative strategy aiming at the reduction of $k$th SBS' contribution to the denominator of $\text{SINR}^{(j)}_{(\sx,i)}$, $\forall j \in [1,K]\setminus{k}$. In this case, the scenario reduces to a point-to-multi-point link between the $k$th SBS and its $K-1$ \linebreak non-served SUEs, whose equivalent channel representation can be expressed as
\begin{IEEEeqnarray}{rCl} \label{eq:Tss_kk}
\Tm_{\sx\sx}^{(k,[k])} &=& [\Tm_{\sx\sx}^{(k,1)\text{T}}, \dots, \Tm_{\sx\sx}^{(k,k-1)\text{T}}, \Tm_{\sx\sx}^{(k,k+1)\text{T}}, \dots, \Tm_{\sx\sx}^{(k,K)\text{T}}]^{\text{T}} \nonumber \\ \: &=& [\tv_{\sx\sx,1}^{(k,[k])} \hspace{1.8mm}| \hspace{1mm} \dots \hspace{1mm} | \hspace{1.8mm} \tv_{\sx\sx,L}^{(k,[k])}]\in \mathbb{C}^{N(K-1) \times L}.
\end{IEEEeqnarray}
The impact of the $j$th SBSs' transmission on \eqref{eq:SINR_ki}, $\forall j \neq k$, is not known at the $k$th SBS. Thus, without loss of generality, the latter can safely assume that
\begin{equation} \label{eq:t^jj_eq_ab}
\begin{cases}
\tv_{\sx\sx,i}^{(j,j)\text{H}}\tv_{\sx\sx,i}^{(j,j)}&= 1, \hspace{4mm} \forall j \hspace{1.5mm} \in [1,K]\setminus{k},  \\
\tv_{\sx\sx,i}^{(m,j)\text{H}}\tv_{\sx\sx,i}^{(m,j)}&= 0, \hspace{4mm} \forall m \in [1,K]\setminus{\{j,k\}}
\end{cases},
\end{equation}
and compute an approximation of \eqref{eq:SINR_ki} accordingly. Note that, in \eqref{eq:t^jj_eq_ab}, the first approximation is related to the direct link between the $j$th SBS/SUE pair, and the second to the co-tier interference generated by the $j$th SBS towards its non-served SUEs. Consequently, the $k$th SBS derives $\widehat{R}_{\sx}$, approximation of $R_{\sx}=\sum_{i=1}^K R_{\sx}^{(i)}$, overall spectral efficiency of the second tier, as
\begin{equation} \label{eq:spec_eff_R3}
\widehat{R}_{\sx} = \frac{1}{N+L}\sum_{i=1}^{L} \log_2{\left(1 + \frac{1}{\sigma^2 + \tv_{\sx\sx,i}^{(k,[k])\text{H}}\tv_{\sx\sx,i}^{(k,[k])}}\right)},
\end{equation}
where, as before, a uniform power allocation at the SBSs is assumed. Let  $\gv_\Tm = (\tv_{\sx\sx,1}^{(k,[k])\text{H}}\tv_{\sx\sx,1}^{(k,[k])},\dots,\tv_{\sx\sx,L}^{(k,[k])\text{H}}\tv_{\sx\sx,L}^{(k,[k])})$ be the vector containing the power values associated to the $L$ columns of $\Tm_{\sx\sx}^{(k,[k])}$. Now let $\gv_{\Tm}^{\theta} \in [1,L]$ be the $\theta$-sized vector containing the indices of the $\theta$ smallest elements of $\gv_{\Tm}$, and $\ev_{i}$ be the vector of the canonical basis with its $i$th entry equal to 1. At this stage, we can define $\Theta^{(k)} = [\ev_{\gv_{\Tm}^{\theta}(1)}|\dots,|,\ev_{\gv_{\Tm}^{\theta}(\theta)}]$ as the outer precoder that selects the columns of $\Tm_{\sx\sx}^{(k,[k])}$ that minimize the co-tier interference generated by the $k$th SBS towards its non-served SUEs. Thus, the input signal at the $k$th SBS is aligned to the most efficient subspace of $\spanV{(\Em^{(k)})}$ for co-tier interference mitigation, and the overall spectral efficiency of the second tier in \eqref{eq:spec_eff_R3} can be rewritten as 
\begin{equation}
\widehat{R}_{\sx}(\theta)~=~\frac{1}{N+L}\sum_{i\in \gv_\Tm^{\theta}} \log_2{\left(1 + \frac{1}{\sigma^2 + \tv_{\sx\sx,i}^{(k,[k])\text{H}}\tv_{\sx\sx,i}^{(k,[k])}}\right)}.
\end{equation}

The choice of $\theta$ is the second degree of freedom available at the $k$th SBS to design $\Thetam^{(k)}$. As previously seen, $\theta$ has a direct impact on the diversity effect provided by the equivalent channel matrices. Intuitively, we expect that a bigger $K$, number of SBS/SUE pairs, will yield a smaller value for $\theta$ to maximize the spectral efficiency of the $k$th SBS/SUE pair, and vice versa. Consequently, from now on we will refer to $\theta(K)$ to explicitly show this dependency. Due to the \linebreak self-organizing nature of the second tier, and to the aforementioned lack of knowledge on the eigenvalues distribution of finite dimension Toeplitz matrices and respective kernels, no analytic optimization of the parameter can be performed. A numerical approach is the only viable way to identify a suitable spectral efficiency maximizing $\theta(K)$ at the $k$th SBS. Assuming a given model for the cross- and co-tier channels, e.g., the channel models described in Sec. \ref{sec:model}, each SBS can find a numerical solution by means of offline Monte Carlo simulations, iterating the following algorithm until statistical relevance of the result is reached.
\begin{algorithm}[h]
\caption{Optimal $\theta(K)$}
	\begin{algorithmic}[1]
	\REQUIRE Set a value for $K$ and select an approach between CIA A and CIA B 
	\FOR{$\theta$ in $[1,L]$}
	    \STATE Compute $\widehat{R}_{\sx}(\theta)$
	\ENDFOR
	\RETURN $\displaystyle \theta(K)=\argmax_{\theta} \widehat{R}_{\sx}(\theta)$
	\end{algorithmic}
\end{algorithm}

The outcome of this algorithm is a map adoptable by the SBSs to realize the aforementioned implicit coordination mechanism as follows. This map hinges on the channel model peculiar to the surrounding operative environment and can be computed offline as a self-optimizing operation to be performed once, before the transmit operations. No online adjustment to the map is required after the self-optimization, hence no particular timing constraints are imposed on the SBSs for this process. As a design policy, each SBS assumes that the number of SUEs inside its coverage area, corresponds to an equal number of potential neighboring interferers, i.e., other SBSs serving the detected SUEs, that is $K-1$. Note that, in a TDD scenario, the number of surrounding SUEs is given by the number of detected sounding reference signals \cite{rpt:3gpp25.814} provided by each SUE to its serving SBS for channel estimation purposes. This information is used during the transmit procedures to identify the best value for $\theta(K)$ instantaneously, by means of the offline-computed map. As a consequence, the CIA cascaded precoder can be immediately derived, according to the chosen one-shot co-tier interference management strategy. The last step before engaging in the transmission is the choice of the optimal precoder realization and power loading strategy at the $k$th SBS, in the sense of spectral efficiency maximization.

\section{Optimal precoder} \label{sec:precoder_c}

As seen in Sec. \ref{sec:cross_tier}, CIA preserves the spectral efficiency of the primary OFDMA transmission. On the other hand, the spectral efficiency of the self-organizing second tier highly depends on the realization of the precoder $\Zm^{(k)}$ at each SBS. Accordingly, each SBS must self-optimize the spectral efficiency of its link to the served SUE, while complying with the adopted interference management strategy, i.e., CIA A or B. In the following, we will focus on the $k$th SBS/SUE pair. We start from a definition.
\begin{definition}[Semi-unitary precoder]  \label{def:semi}
A precoder \linebreak $\Wm \in \mathbb{C}^{N \times M}$ is \textit{semi-unitary} if and only if $\rank{(\Wm)}=min\{N,M\}$ and all its non-zero eigenvalues are equal to 1, thus $\Wm\Wm\herm=\Id_N$ or $\Wm\herm\Wm=\Id_M$.
\end{definition}
As a consequence, the following holds.
\begin{proposition}[Product of semi-unitary precoders] \label{prop:prod}
The product of $K$ semi-unitary precoders is a semi-unitary precoder. 
\end{proposition}
\begin{thmproof}
See Appendix \ref{sec:appendix_A}.
\end{thmproof}

The following result provides the optimal linear precoder based strategy to be adopted by any transmitter that aims at maximizing the spectral efficiency of its transmission towards a secondary receiver, while fulfilling a feasible interference cancellation constraint w.r.t. the interference link towards a primary receiver.
\begin{proposition}[Optimal interference cancellation precoder] \label{proposition:general}
Consider an interference channel with a primary and a cognitive secondary transmitter/receiver pair, i.e, TX1/RX1 and TX2/RX2 respectively, characterized by the following equations
\begin{equation}
\begin{array}{lcl}
\yv_1 &= \Hm_{11} \xv_1 + \Hm_{21} \xv_2 + \nv_1 \\
\yv_2 &= \Hm_{22} \xv_2 + \Hm_{12} \xv_1 + \nv_2,
\end{array}
\end{equation}
with $\yv_i$ $N$-sized received vectors, $\Hm_{ij} \in \mathbb{C}^{N \times M}$ channel matrices with $N < M$, $\nv_i\sim\Cc\Nc(0, \sigma^2\Id_N)$ AWGN vector and $\xv_i=(\xv_{i,j},\dots,\xv_{i,M})^{\text{T}} \in \mathbb{C}^{M \times 1}$ transmit vectors. When perfect CSI w.r.t. $\Hm_{21}$ is available at TX2, the interference towards RX1 can be canceled by means of a linear null-space precoder. If also perfect CSI w.r.t. $\Hm_{22}$ is available, a \textbf{semi-unitary} precoder is optimal in the sense of the spectral efficiency of the secondary link under the interference cancellation constraint.
\end{proposition}

\begin{thmproof}
See Appendix \ref{sec:appendix_B}.
\end{thmproof}

Note that, Proposition \ref{proposition:general} holds true for any configuration of the interference channel, as long as the system is characterized by the equations provided in the hypothesis. In particular, the result is independent from parameters such as bandwidth, number of antennas, number of subcarriers and so on. Consequently, we can state the following result.

\begin{corollary}[Optimal CIA precoder] \label{corol:cia}
Consider a two-tiered network where a licensee OFDMA base station coexists with several non-cooperative single antenna opportunistic base stations adopting CIA. The cascaded CIA precoder $\Zm^{(k)}$ is optimal, in the sense of maximum link spectral efficiency for the $k$th SBS/SUE pair, regardless of the chosen co-tier interference mitigation approach.
\end{corollary}

\begin{thmproof}
See Appendix \ref{sec:appendix_C}.
\end{thmproof}

\section{Spectral efficiency computation} \label{sec:spec}

At this stage, we dispose of all the necessary tools to derive the spectral efficiency of the two tiers. We recall that, in \linebreak Sec. \ref{sec:co_tier}, the cross-tier interference generated by the MBS towards the SUEs has been omitted for simplicity from the equations describing CIA. Even though not necessary at the $k$th SBS to derive $\Thetam^{(k)}$, this interference may strongly limit the achievable spectral efficiency at the $k$th SUE. Accordingly, hereafter the cross-tier interference generated by the MBS will be taken into account in the performance evaluation of the considered two-tiered network. We remark that, \linebreak macro-cell and small-cell coverage areas have very different size, thus the distance between the MBS and the SUEs served by the active SBSs hinges on the second tier deployment and varies depending on the considered network layout. In order to capture this crucial parameter, we define an interference factor $\alpha \in [0,1]$ to scale the cross-tier interference generated by the MBS towards the $k$th SUE. In particular, $\alpha \simeq 1$ models a scenario where the active SBSs are operating nearby the MBS, whereas if the second tier is deployed very far the MBS we have $\alpha \simeq 0$. Moreover, for the sake of compactness of the notation, we let
\begin{equation}
\begin{array}{lclcl} \overline{\Tm}_{\sx\px}^{(k,j)} =\tilde{\Tm}_{\sx\px}^{(k,j)} \Thetam^{(k)} = [\overline{\tv}_{\sx\px,1}^{(k,j)} \hspace{1.8mm}| \hspace{1mm} \dots \hspace{1mm} | \hspace{1.8mm} \overline{\tv}_{\sx\px,N}^{(k,j)}]^{\text{T}} \in \mathbb{C}^{N \times \theta} \\
\overline{\Tm}_{\sx\sx}^{(k,k)} =\Tm_{\sx\sx}^{(k,k)} \Thetam^{(k)} = [\overline{\tv}_{\sx\sx,1}^{(k,k)} \hspace{1.8mm}| \hspace{1mm} \dots \hspace{1mm} | \hspace{1.8mm} \overline{\tv}_{\sx\sx,N}^{(k,k)}]^{\text{T}} \in \mathbb{C}^{N \times \theta} \end{array}
\end{equation}
be the equivalent channel representation of the link connecting the $k$th SBS to the $j$th MUE and $k$th SUE, respectively. The spectral efficiency for perfect and imperfect CSI is computed as follows.

\subsection{Perfect CSI} 
We start by noting that for perfect CSI at the SBSs, the spectral efficiency at the $j$th MUE is given by \eqref{eq:spec_eff_Rp}. Therefore, we can switch our focus to the spectral efficiency at the $k$th SUE. Let
\begin{IEEEeqnarray}{rCl}
I_{(\sx, i)}^{(k)} &=& \sum_{m \neq k}^K \overline{\tv}_{\sx\sx,i}^{(m,k)\text{H}} d(p^{(m)}_{\sx,1},\dots,p^{(m)}_{\sx,\theta}) \overline{\tv}_{\sx\sx,i}^{(m,k)} + ... \nonumber \\ && +\: \alpha p_{\px,i}\left|\left[\Fm \Bm \Hm_{\px\sx}^{(1,k)}\Am\Fm^{-1}\right]_{i,i}\right|^2 + \sigma_n^2
\end{IEEEeqnarray}
be the interference plus noise component of the $i$th received symbol at the $k$th SUE, with $\alpha p_{\px,i}\left|\left[\Fm \Bm \Hm_{\px\sx}^{(1,k)}\Am\Fm^{-1}\right]_{i,i}\right|^2$ representing the cross-tier interference generated by the MBS. Then, \eqref{eq:SINR_ki} can be redefined as
\begin{equation} \label{eq:SINR_s_final}
\text{SINR}_{(\sx, i)}^{(k)} = \frac{\overline{\tv}_{\sx\sx,i}^{(k,k)\text{H}} d(p^{(k)}_{\sx,1},\dots,p^{(k)}_{\sx,\theta}) \overline{\tv}_{\sx\sx,i}^{(k,k)}}{I_{(\sx, i)}^{(k)}},
\end{equation}
and the spectral efficiency at the $k$th SUE can be computed by plugging \eqref{eq:SINR_s_final} into \eqref{eq:spec_eff_R}.
 
\subsection{Imperfect CSI} \label{sec:imperf_csi}

In realistic implementations of wireless communications systems, the transmitter only disposes of noisy channel estimations, i.e., imperfect CSI. The design of a channel estimation procedure is out of the scope of this work, hence we assume a classic training/transmission scheme as in \cite{art:hass03} for simplicity. We consider a block fading channel model characterized by a coherence time $\Tc$. Thus, each transmitter must perform the necessary channel estimations with periodicity $\Tc$, in both tiers. We denote as $\tau \leq \Tc$ the time spent estimating the channel. The available time for transmission at each SBS, until the next channel estimation is necessary, is then $\Tc-\tau$. Accordingly, a longer $\tau$ yields better channel estimations, but reduces the available time for transmission. Typically, a channel estimation can be represented as \cite{art:hass03}
\begin{equation*}
\rv = \sqrt{\rho \tau}\hv + \nv,
\end{equation*}
where $\hv$ is the channel vector, $\rho$ is the transmit power and \linebreak $\nv\sim\Cc\Nc(0, \sigma_n^2\Id_{(l+1)})$ is the AWGN at the devices' antennas. We assume that each transmitter estimates $\hv$ by evaluating the observation $\rv$, e.g., by means of a minimum mean-square error (MMSE) approach, to obtain
\begin{equation*}
\hv = \hat{\hv} + \tilde{\hv}, 
\end{equation*}
with $\hat{\hv}$ desired channel estimation and $\tilde{\hv}$ independent error. Note that, in order to derive $\Em^{(k)}$ and fulfill the cross-tier interference constraint in \eqref{eq:nullcondition} at the $k$th SBS, a perfect channel estimation of $\hv_{\sx\px}^{(k,j)}$ is required, $\forall j \in [1,M]$. If perfect CSI is not available, $\overline{\Tm}_{\sx\px}^{(k,j)}\neq0$ and the SBSs generate cross-tier interference towards the MUEs. Consequently, the SINR of $i$th received symbol at the $j$th MUE reads
\begin{equation} \label{eq:SINR_p_imp}
\text{SINR}_{(\px, i)}^{(j)} = \frac{p_{\px,i} \left|\left[\Fm \Bm \Hm_{\px\px}^{(1,j)}\Am\Fm^{-1}\right]_{i,i}\right|^2}{\sum_{k=1}^K \overline{\tv}_{\sx\px,i}^{(k,j)\text{H}} d(p^{(k)}_{\sx,1},\dots,p^{(k)}_{\sx,\theta}) \overline{\tv}_{\sx\px,i}^{(k,j)} + \sigma_n^2}.
\end{equation}
As seen in Sec. \ref{sec:co_tier}, the $k$th SBS adopting CIA A or CIA B needs CSI w.r.t. the links towards the served SUE or neighboring SUEs respectively. We note that, the SINR of the $i$th received symbol at the $k$th SUEs is always computed by \eqref{eq:SINR_s_final}. Nevertheless, imperfect CSI may decrease the effectiveness of the outer precoder $\Thetam^{(k)}$ and yield higher co-tier interference, i.e., the term $\sum_{m \neq k}^K \overline{\tv}_{\sx\sx,i}^{(m,k)\text{H}} d(p^{(m)}_{\sx,1},\dots,p^{(m)}_{\sx,\theta}) \overline{\tv}_{\sx\sx,i}^{(m,k)}$ in \eqref{eq:SINR_s_final}. This in general worsens the SINR per received symbol at the SUEs, resulting in spectral efficiency losses. Additionally, we know from \cite{art:hass03} that the time and resources spent for channel estimation have an impact on the effective SINR experienced at each receiver. If training and data symbols carry the same average power, we can define $\overline{\text{SINR}}^{(j)}_{(\cdot,i)}$, effective SINR value of the $i$th symbol at the $j$th receiver, as    
\begin{equation} \label{eq:SINR_eff_imp}
\overline{\text{SINR}}^{(j)}_{(\cdot,i)}=\frac{\bigg(\text{SINR}^{(j)}_{(\cdot,i)} \bigg)^2 \tau}{1 + (1 + \tau)\text{SINR}^{(j)}_{(\cdot,i)}}, \quad \forall j \in [1,KN].
\end{equation}
Therefore, we can compute $R^{\text{I}}_{\px}$ and $R^{\text{I}}_{\sx}$, spectral efficiency of first and second tier respectively, for imperfect CSI, and obtain 
\begin{eqnarray} 
R^{\text{I}}_{\px} &=& \frac{\Tc-\tau}{\Tc(N+L)}\sum_{j=1}^M \sum_{i=1}^{N} \log_2(1 + \overline{\text{SINR}}^{(j)}_{\px,i}) \label{eq:spec_eff_imp_p} \\
R^{\text{I}}_{\sx} &=& \frac{\Tc -\tau}{\Tc(N+L)} \sum_{m=1}^K \sum_{i=1}^{N} \log_2(1 + \overline{\text{SINR}}^{(m)}_{\sx,i}). \label{eq:spec_eff_imp_s}
\end{eqnarray}

\section{Numerical analysis} \label{sec:num_anal}

In this section we focus on the achievable spectral efficiency of the proposed CIA scheme, for several configurations of the studied two-tiered network. Throughout the analysis, an interference factor $\alpha=1$ is assumed, unless otherwise stated. Extensive Monte Carlo simulations are performed to obtain statistically relevant results. No particular channel model is adopted in this study, hence we consider frequency-selective Rayleigh fading channels with uniform power delay profile, as described in Sec. \ref{sec:model}. The study of the performance of CIA for extended cellular layouts, characterized by multiple macro-cells in the first tier, structured SBSs' positioning in the second tier and practical channel models, will be matter of our future research. For computational tractability, we assume that the OFDMA transmission at the MBS is performed over a bandwidth of $1.92$ MHz, divided in $N=128$ subcarriers, with cyclic prefix size of $L=l=32$, as in the extended mode of the least resource demanding LTE/LTE-A downlink configuration \cite{rpt:3gpp25.814}. First we analyze the performance for perfect CSI in the second tier. We compare our results to what is achievable by means of an orthogonalization strategy such as TDMA, where only one SBS is active at each iteration (time slot), considered an appropriate benchmark for the following reasons. TDMA is the first traditional benchmark for the performance of interference alignment solutions for the interference channel \cite{art:cadambe2008i}. In fact, it is a classical distributed solution for self-organizing and ad-hoc networks \cite{art:rhee09}, commonly adopted in many commercial standards and applications \cite{stan:IEEE09}, when no cooperation or communication can be established between potentially interfering transmitters. Finally, differently from a frequency division multiple access (FDMA) approach, equivalent in terms of achievable spectral efficiency, a TDMA approach to address the co-tier interference issue in the second tier is compliant with the frequency reuse 1 assumption made in Sec. \ref{sec:model}. As a second step, we investigate the robustness of CIA when dealing with channel estimation errors, i.e., imperfect CSI. Afterwards, we complete our study by analyzing the percent increase in achievable spectral efficiency that the proposed approach could yield w.r.t. the legacy single tier based network deployment. Accordingly, we compare the spectral efficiency of a \linebreak two-tiered network composed of an OFDMA MBS underlaid with $K$ self-organizing SBSs adopting CIA, with the performance of a standalone legacy OFDMA MBS, for imperfect CSI and different values of $\alpha$.

\begin{figure}[!h]
	\centering
	\includegraphics[width=.99\columnwidth]{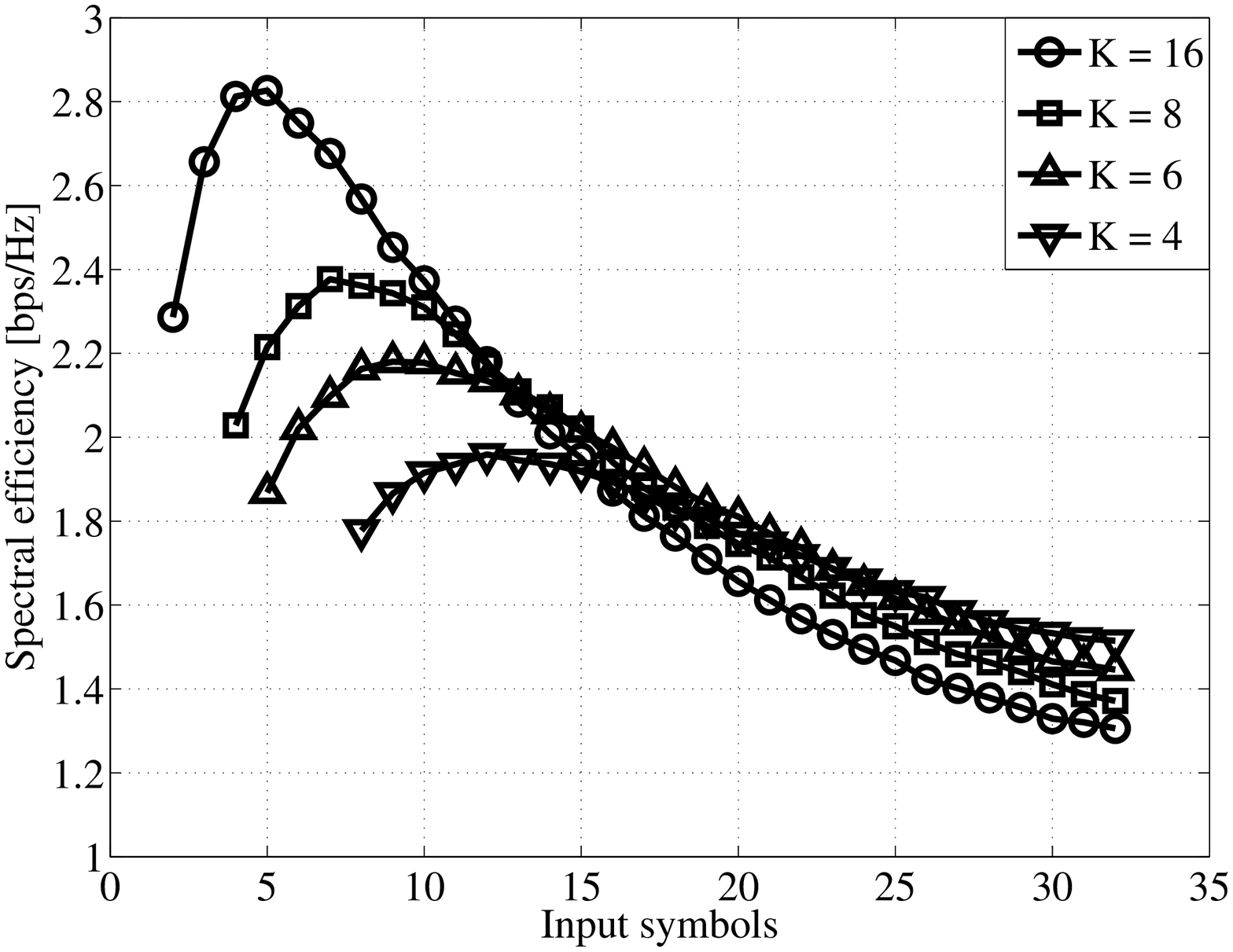}
	\caption{Spectral efficiency of the second tier for CIA A as the dimension of the input signal subspace changes, for $K \in \{4,6,8,16\}$ SBSs. $N=128$, $L=32$ and bandwidth of 1.92~Mhz.}	
	\label{fig:d_max}
\end{figure}

\begin{figure}[h]
	\centering
	\includegraphics[width=.99\columnwidth]{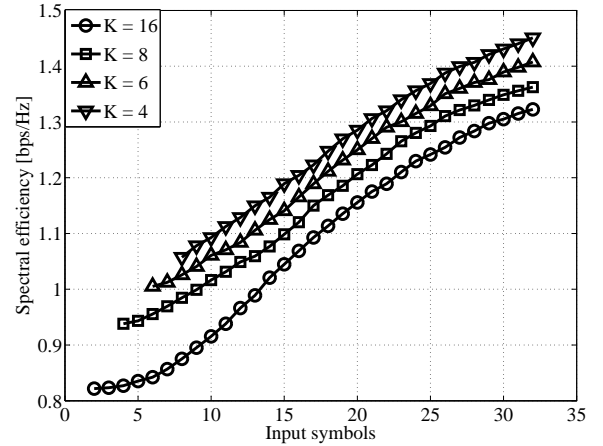} 
	\caption{Spectral efficiency of the second tier for CIA B as the dimension of the input signal subspace changes, for $K \in \{4,6,8,16\}$ SBSs. $N=128$, $L=32$ and bandwidth of 1.92~Mhz.}
	\label{fig:d_min}
\end{figure}

As a preliminary study, we identify the optimal $\theta(K)$, for both CIA A and CIA B, iterating the algorithm described in Sec. \ref{sec:co_tier}. We aim at finding the optimal input signal subspace dimension at the SBSs such that the second tier does not suffer from co-tier interference limitation. The considered thermal noise at each SUE is such that 
\begin{equation} \label{eq:SNR_numerical}
SNR^{(k)}_{(\sx,i)} = \mathbb{E}\bigg[\log_{10}\left(\frac{\tv_{\sx\sx,i}^{(k,k)\text{H}}\tv_{\sx\sx,i}^{(k,k)}}{\sigma^2}\right)\bigg] = 30 \text{dB},
\end{equation}
$\forall i \in [1,N], k \in [1,K]$. We let $K \in \{4,6,8,16\}$ and depict the performance for CIA A and CIA B in Fig. \ref{fig:d_max} and Fig. \ref{fig:d_min}, respectively. 
For CIA A, a clear dependency of the optimal input signal subspace dimension on the number of active SBSs is shown, further motivating the intuition given in Sec. \ref{sec:co_tier}, i.e., the larger $K$ the smaller the optimal $\theta(K)$. Intuitively, we would expect that the co-tier interference increases with the number of active users, regardless of the chosen approach. Conversely, the achievable spectral efficiency in case of optimal $\theta(K)$ increases with $K$ thanks to an effective co-tier interference mitigation, and a consistent robustness against the cross-tier interference generated by the MBS. This is not verified for CIA B, where the performance of the second tier is highly interference limited and the optimal $\theta(K)$ is always equal to $L$, independently of $K$. As a result, CIA B is not sufficient to mitigate the co-tier interference in the second tier, i.e., the higher $K$ the lower the achievable spectral efficiency. Therefore, an adequate input signal subspace reduction at the $k$th SBS, followed by a selfish maximization of the spectral efficiency of the link towards the served SUE (CIA A), provides a higher receive diversity resulting in a significant power gain at the $k$th SUE. 

\begin{figure}[h]
	\centering
	\includegraphics[width=.99\columnwidth]{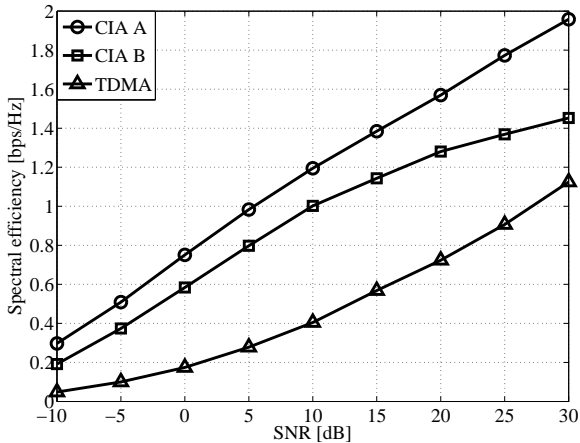}
	\caption{Spectral efficiency of the second tier as the SNR changes, $K=4$ SBSs, $N=128, L=32$ and bandwidth of 1.92~Mhz.}
	\label{fig:4_users}
\end{figure}

\begin{figure}[h]
	\centering
	\includegraphics[width=.99\columnwidth]{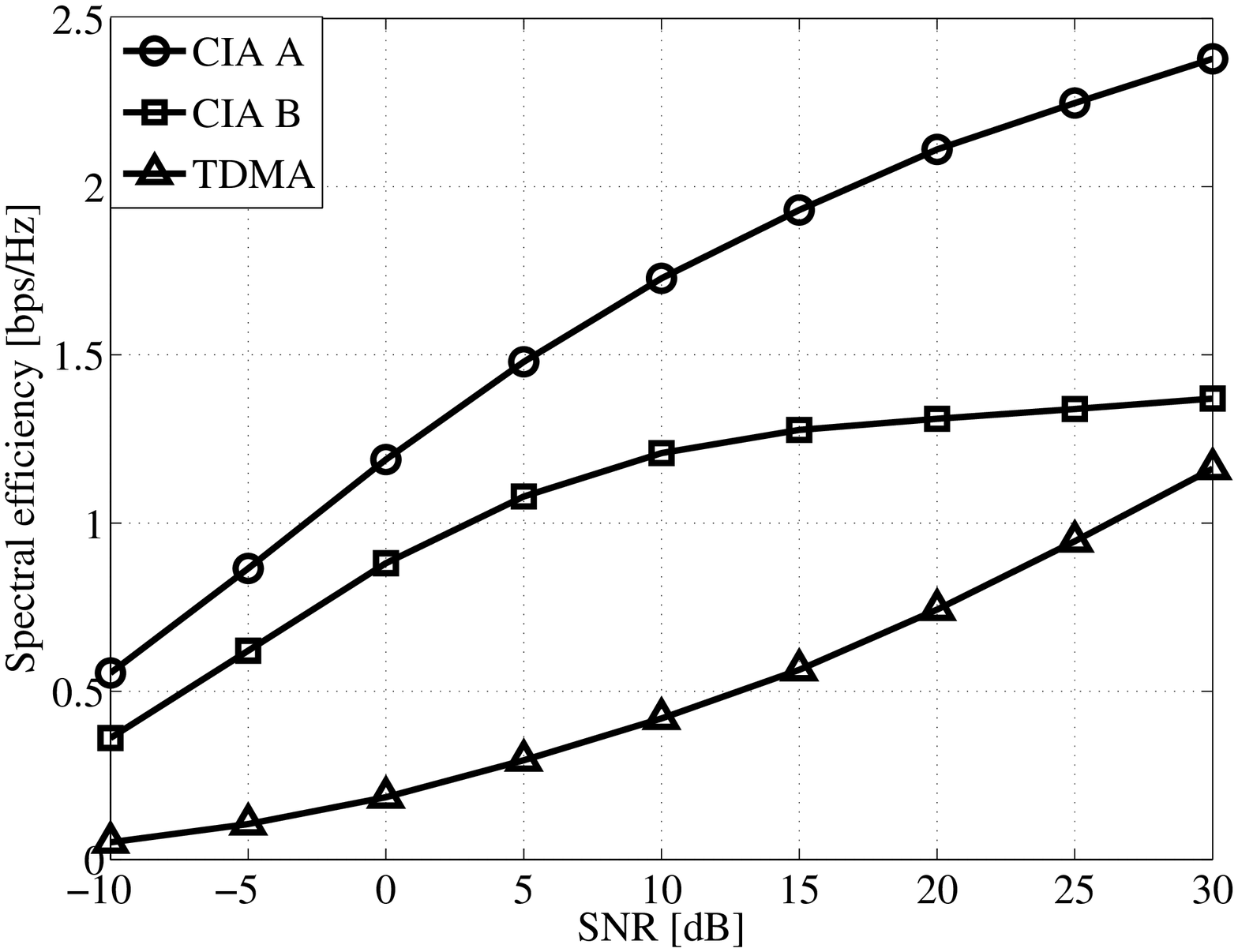}
	\caption{Spectral efficiency of the second tier as the SNR changes, $K=8$ SBSs, $N=~128, L=32$ and bandwidth of 1.92~Mhz.}
	\label{fig:8_users}
\end{figure}

\begin{figure}[h]
	\centering
	\includegraphics[width=.99\columnwidth]{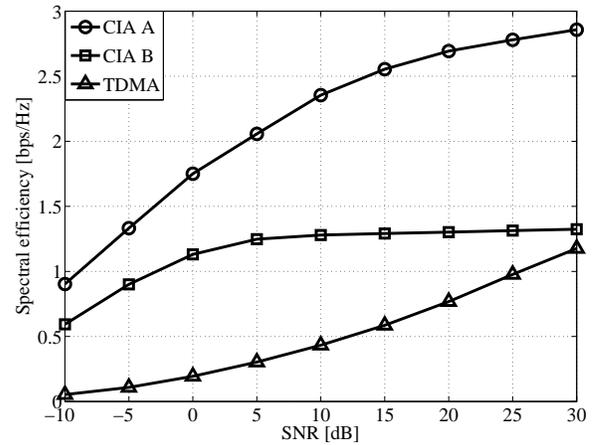}
	\caption{Spectral efficiency of the second tier as the SNR changes, $K=16$ SBSs, $N=128, L=32$ and bandwidth of 1.92~Mhz.}
	\label{fig:16_users}
\end{figure}

Now we consider the previously obtained optimal $\theta(K)$ and let $SNR^{(k)}_{(\sx,i)}~\in~[-10,30]$~dB. We compute the achievable spectral efficiency for the two proposed methods for $K\in\{4, 8, 16\}$, in Fig. \ref{fig:4_users}, Fig. \ref{fig:8_users} and Fig. \ref{fig:16_users}, respectively. 
Significant SNR gains brought by CIA A over CIA B are evident for each configuration, confirming the previous findings. In particular, the larger $K$ the higher the experienced SNR gain. As previously stated, we provide a comparison with a commonly implemented distributed TDMA approach, where only one SBS is active at each iteration (time slot). We remark that, to guarantee a fair comparison, the active SBS in the TDMA scheme adopts an optimal water-filling power loading strategy \cite{book:tse2005f}. Interestingly, both proposed methods outperform the TDMA approach in the considered SNR range. Furthermore, if we focus on the best performer out of the proposed strategies, i.e., CIA A, we observe remarkable SNR gains up to $20$~dB for $K=4$ and $30$~dB for $K=16$ w.r.t. TDMA. We note that, an SBS adopting CIA A requires CSI w.r.t. a lower number of channels, i.e., the direct link towards its served SUE, whereas for CIA B the CSI related to all the links towards the $K$ SUEs inside its coverage area are needed. Thus, CIA A has the same CSI requirements as TDMA and is not affected by scalability issues as $K$ increases. Additionally, CIA A is more robust than CIA B to the cross-tier interference generated by the MBS. In fact, it yields an always increasing achievable spectral efficiency for the considered values of $K$ and SNR range, hence is a preferable choice in the sense of the overall self-organizing second tier performance. For these reasons, in the remainder of the section, we will consider CIA A as the selected approach to implement CIA.

\begin{figure}[h]
	\centering
	\includegraphics[width=.99\columnwidth]{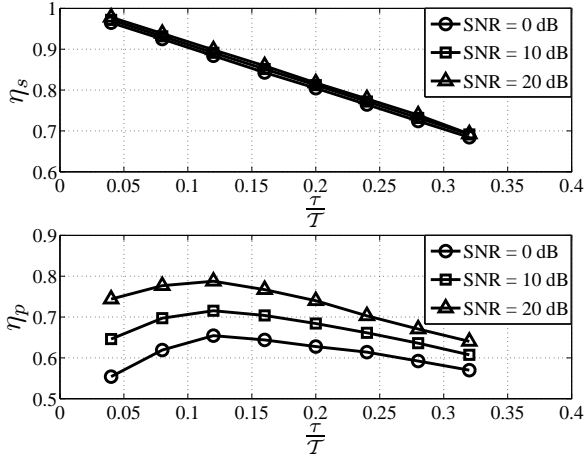}
    \caption{Ratio between the achievable spectral efficiency of the SBSs and MBS for imperfect CSIT and perfect CSIT in the second tier, $K=8$ SBSs, SNR $=\{0,10,20\}$ dB, $N=128, L=32$ and bandwidth of 1.92~Mhz.}
	\label{fig:eta_k4}
\end{figure}

Now we compute the performance of CIA when perfect CSI is not available in the second tier as discussed in \linebreak Sec. \ref{sec:imperf_csi}, and evaluate the impact of an imperfectly built $\Zm^{(k)}$ precoder on the overall network performance as the ratio $\frac{\tau}{\Tc}$ changes. Let $\displaystyle \eta_p=\frac{R^{\text{I}}_{\px}}{R_{\px}}$ and $\displaystyle \eta_s=\frac{R^{\text{I}}_{\sx}}{R_{\sx}}$ be two parameters that measure the percentage of the achievable spectral efficiency for perfect CSI that is achievable when only imperfect CSI is available, for the first and second tier respectively. Consider a first tier with $M=4$ MUEs and a second tier composed of $K=8$ SBS/SUE pairs. In Fig. \ref{fig:eta_k4}, $\eta_s$ and $\eta_p$ are computed as different $\frac{\tau}{\Tc}$ proportions are chosen, for SNR $\in \{0, 10, 20\}$~dB.
We note that, $\eta_s$ can assume values very close to 1 and is inversely proportional to $\frac{\tau}{\Tc}$. In particular, when $\frac{\tau}{\Tc}$ is too high, the time spent to acquire the CSI at the SBSs is such that the experienced SINR gains are not sufficient to compensate the loss induced by the lack of time available for the transmission. In other words, the pre-log factor in \eqref{eq:spec_eff_imp_s} is dominant and $\eta_s$ scales linearly with $\frac{\tau}{\Tc}$. Remarkably, this behavior is independent from the SNR value at the receiver, showing the robustness of the proposed co-tier interference management mechanism to channel estimation errors. 

Furthermore, we recall that each SBS makes use of only the CSI related to the direct link towards its served SUE to compute the CIA A precoder. Thus, the spectral efficiency loss for imperfect CSI w.r.t. the perfect CSI case, measured by $\eta_s$, is not due to additional co-tier interference generated towards the non-served SUEs, but only to a less effective power allocation w.r.t. the direct link connecting each SBS/SUE pair. As a consequence, a larger number of active SBSs, i.e. $K$, would not yield different values for $\eta_s$. This demonstrates the scalability of the proposed solution. Switching our focus to $\eta_p$ we clearly see a constructive impact of a longer $\tau$ on the effectiveness of the cross-tier interference alignment technique, thus on the spectral efficiency of the first tier. In particular, differently from the previous case, the pre-log factor in \eqref{eq:spec_eff_imp_p} becomes dominant ($\eta_p$ scales linearly with $\frac{\tau}{\Tc}$) only for $\frac{\tau}{\Tc}>0.12$. In fact, by estimating the channels during the optimal $\frac{\tau}{\Tc}$ portion of the coherence time, i.e., $\frac{\tau}{\Tc}=0.12$, regardless of the power of the noise affecting the estimations, each SBS can compute a more precise precoder $\Em^{(k)}$, inducing a consequent power gain at the MUEs. We note that, in general, the cross-tier interference alignment strategy is not as robust as the co-tier interference management mechanism against imperfect CSI. For very low SNR values, i.e., $0$~dB, the first tier's loss is around $36\%$ of its achievable spectral efficiency for perfect CSI, whereas for medium to high SNR, i.e., $20$~dB, the loss can be reduced to $21\%$. 

\begin{figure}[h]
	\centering
	\includegraphics[width=.99\columnwidth]{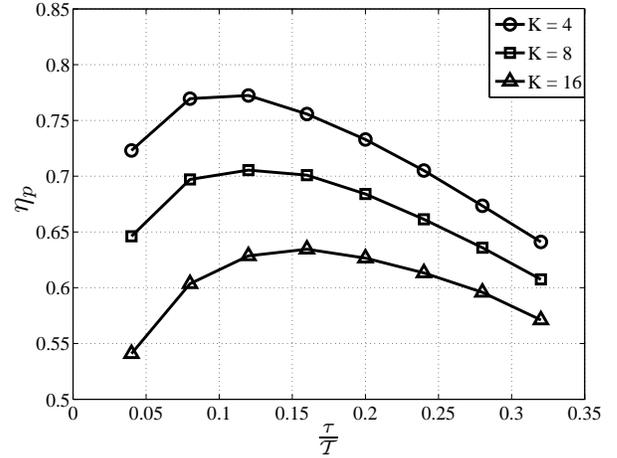}
	\caption{Ratio between the achievable spectral efficiency of the MBS for imperfect CSIT and perfect CSIT in the second tier, $K=\{4,8,16\}$ SBSs, SNR $=10$ dB, $N=128, L=32$ and bandwidth of 1.92~MHz.}
	\label{fig:etap_k468}
\end{figure}

We keep focusing on the first tier and analyze the impact of the cross-tier interference generated by the second tier as the number of active SBSs increases. In Fig. \ref{fig:etap_k468}, $\eta_p$ is computed for $M=4$ MUEs and $K \in \{4,8,16\}$ SBSs.
We note that, to achieve the target average spectral efficiency per macro-cell in LTE-A, ranging between 2.4 and 3.7 bit/s/Hz \cite{rpt:3gpp36.913}, a target average SNR ranging between 9 and 11 dB is required, for practical modulation and coding schemes \cite{rpt:3gpp36.942}. Accordingly, in this analysis we assume SNR$=10$ dB, i.e., a likely occurring value in future LTE-A networks to be able to meet the target performance. As could be intuitively expected, a direct proportionality of the optimal $\frac{\tau}{\Tc}$ and $K$ is evident, e.g., the optimal $\frac{\tau}{\Tc}=0.16$ for $K=16$ is higher than the optimal $\frac{\tau}{\Tc}$ for $K=\{4,8\}$. Furthermore, the larger $K$ the higher the loss experienced by the first tier, due to the superior cross-tier interference generated by the second tier. In particular, the spectral efficiency loss when $K=16$ is around $15\%$ more than what is obtained for $K=4$. These results confirm our previous findings. The sensitivity of CIA to channel estimation errors in terms of first tier's loss is confirmed. However, an optimal $\frac{\tau}{\Tc}$ can be found even if the number of SBSs increases, improving the effectiveness of the cross-tier interference alignment. 

\begin{figure}[th!]
	\centering
	\includegraphics[width=.99\columnwidth]{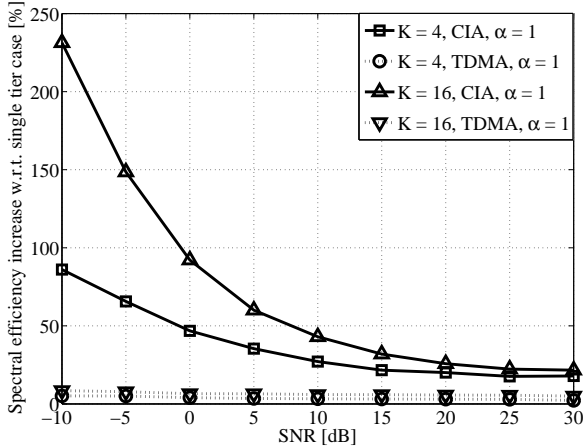} 
	\caption{Percent increase in spectral efficiency w.r.t. the OFDMA-based single tier case. $K \in \{4,16\}$, $N=128, L=32$ and bandwidth of 1.92~MHz. Full cross-tier interference from the MBS to the SUEs.}
	\label{fig:sum_se}
\end{figure}

\begin{figure}[h]
	\centering
	\includegraphics[width=.99\columnwidth]{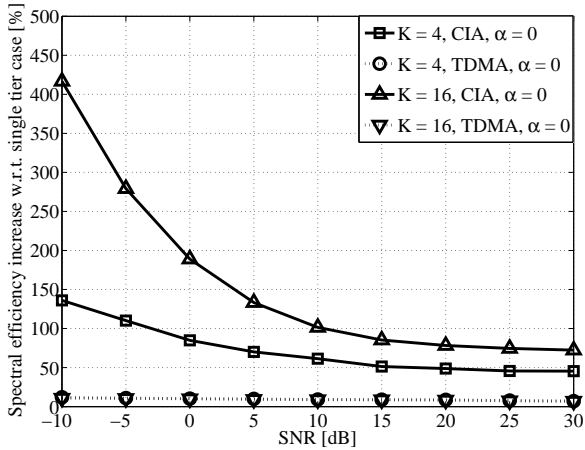}
	\caption{Percent increase in spectral efficiency w.r.t. the OFDMA-based single tier case. $K \in \{4,16\}$, $N=128, L=32$ and bandwidth of 1.92~MHz. No cross-tier interference from the MBS to the SUEs.}
	\label{fig:sum_se_noint}
\end{figure}

As a final step, we study the performance of the two-tiered network, to evaluate the advantages, if any, brought by the proposed technique. We consider as a reference performance the achievable spectral efficiency of a standalone OFDMA MBS serving $M=4$ MUEs, by means of an optimal power allocation strategy given by a classic water-filling algorithm \cite{book:tse2005f}. The second tier is composed of $K \in \{4,16\}$ SBS/SUE pairs, and given the previous results, we assume $\frac{\tau}{\Tc}=0.12$. For clarity, we compute the percent increase in achievable spectral efficiency brought to the single standalone MBS (reference performance) by the co-channel deployment of a second tier of self-organizing SBSs adopting CIA. Note that, for completeness, the achievable percent increase in case of TDMA approach in the second tier is computed as well. As can be seen in Fig. \ref{fig:sum_se}, a second tier adopting CIA provides significant additional spectral efficiency to the legacy single tier performance at any SNR regime. 
Remarkably, CIA outperforms the TDMA approach for both values of $K$. Interestingly, the lower the SNR the larger the advantage of CIA w.r.t. TDMA, showing the effectiveness of CIA for practically relevant SNR values. We note that, at very high SNR an increase in the number of SBSs from $K=4$ to $K=16$ does not yield a significant advantage due to the cross-tier interference coming from the MBS. Conversely, for SNR$=0$ dB, increasing the number of SBSs from $K=4$ to $K=16$ doubles the percent increase (from $45\%$ to $90\%$), whereas for SNR$=10$ dB, the percent increase goes from $27\%$ to $42\%$ (more than $50\%$ of relative increase). Similar insights can be drawn from Fig. \ref{fig:sum_se_noint}, where the analysis is repeated for $\alpha=0$, i.e., no cross-tier interference generated by the MBS towards the SUEs.
In particular, for SNR$=0$ dB, increasing the number of SBSs from $K=4$ to $K=16$ more than doubles the percent increase w.r.t. the single tier approach (from $90\%$ to $190\%$), whereas for SNR$=10$ dB, the percent increase goes from $60\%$ to $100\%$ ($60\%$ of relative increase). On the other hand, in this case the deployment of $K=16$ provides around $70\%$ of percent increase for very high SNR, i.e. SNR$=30$ dB, whereas the result for $K=4$ is around $45\%$ (more than $50\%$ of relative increase). These results imply that the robustness and consistency of CIA is such that the achievable spectral efficiency in the second tier compensates the loss experienced in the first tier due to the imperfect computation of $\Zm^{(k)}$. This insight can be drawn $\forall \alpha \in [0,1]$, particularly from very low to medium SNR regime. Thus, additional capacity can always be added to the network, regardless of the distance between the MBS and the SUEs.

\section{Conclusion} \label{sec:conclusion}

In this work, we have addressed the problem of the coexistence of multiple cells inside a two-tiered network, where a primary oblivious OFDMA MBS and an opportunistic \linebreak self-organizing tier of SBSs perform a non-cooperative downlink transmission towards a group of MUEs/SUEs respectively, with frequency reuse 1. A novel cognitive interference alignment (CIA) scheme has been proposed to manage the resulting cross- and co-tier interference, yielding spectral efficiency enhancements for the overall network. CIA nulls the \linebreak cross-tier interference and preserves the degrees of freedom of the legacy OFDMA transmission regardless of the transmit power at the SBSs. Thus, the presence of the legacy interference free dimensions at the MUEs is guaranteed, and additional transmit dimensions to each SBS are provided. As a second step, the co-tier interference in the secondary system is mitigated by means of an input signal subspace reduction self-organizing strategy. The optimal precoder that maximizes the spectral efficiency of the link connecting each SBS to its served SUE is found through a distributed \linebreak one-shot strategy. Our numerical findings show significant spectral efficiency improvements with respect to legacy \linebreak TDMA/FDMA approaches as the number of self-organizing SBSs increases. The design of the proposed linear cascaded precoder only requires that each SBS is aware of the number of SUEs inside its coverage area, and disposes of a perfect local CSI w.r.t. the link towards the served SUEs and the MUEs reached by its transmission. Remarkably, CIA exhibits consistent robustness against channel estimation errors, yielding promising spectral efficiency results. Finally, we evaluate the percent increase in spectral efficiency that a two-tiered network deployed by means of CIA experiences w.r.t, an OFDMA MBS, regardless of the presence of channel estimation errors in the second tier. We show that, CIA yields a performance enhancement at any SNR regime, despite the non-negligible cross-tier interference generated by the MBS towards the second tier. This work is the first step towards the characterization of the performance of a self-organizing two-tiered network, where the self-organization is realized at physical layer, without requiring signaling or cooperation between the transmitters. Accordingly, it is a matter of our future research the adaptation of CIA to cellular layouts including multiple MBSs, structured SBSs' positioning and practical channel models. 

\appendices
\section{Proofs}
\subsection{Proof of Proposition \ref{prop:prod}} \label{sec:appendix_A}
\begin{IEEEproof}
Let $\Wc=\{\Wm_1, \dots, \Wm_K \}$ be a set of $K$ \linebreak semi-unitary precoders $\Wm_i~\in~\mathbb{C}^{N_i \times M_i}$,  $\forall i \in [1,K]$, such that $M_i=N_{i+1}, \forall i \in [1,K-1]$. We know that if $N_i > M_i$ then $\Wm_i\herm \Wm_i = \Id_{M_i}$, whereas if $M_i > N_i$ then $\Wm_i\Wm_i\herm = \Id_{M_i}$. In the following, we will assume $N_i > M_i$. Let $\Qm =~\prod_i^K \Wm_i \in~\mathbb{C}^{N_1 \times M_K}$ be the product of the $K$ precoders. If we compute the matrix given by $\Qm\herm \Qm$, we obtain
\begin{eqnarray*}
\Qm\herm \Qm &=& \Wm_K\herm \dots \Wm_2\herm \Wm_1\herm \Wm_1 \Wm_2 \dots \Wm_K \\
             &=& \Wm_K\herm \dots \Wm_2\herm \Id_{M_1} \Wm_2 \dots \Wm_K \\
             &=& \Wm_K\herm \Id_{M_{K-1}} \Wm_K \\
             &=& \Id_{M_K},
\end{eqnarray*}
where we recursively used the definition of the semi-unitary $\Wm_i$ given in Def. \ref{def:semi}. Consequently, $\Qm$ is semi-unitary. Note that, if $N_i < M_i$, $\Qm \Qm \herm = \Id_{N_K}$ is obtained similarly, demonstrating that $\Qm$ is semi-unitary regardless of the considered case, and this ends the proof.
\end{IEEEproof}

\subsection{Proof of Proposition \ref{proposition:general}} \label{sec:appendix_B}
\begin{IEEEproof}
We start by isolating the interference plus noise component of the received message at RX2, that performs single-user decoding, as $\xiv_2 =  \Hm_{12} \xv_1 + \nv_2$. Analogously, can define $\Sm_{(2,\xi)}$, covariance matrix of $\xiv_2$, as 
\begin{equation} \label{eq:S_eta}
\Sm_{(2,\xi)} =  \Hm_{12} d(p_{1,1},\dots,p_{1,M}) \Hm_{12}\herm + \sigma_n^2 \Id_N,
\end{equation}
where $d(p_{1,1},\dots,p_{1,M})$ is a generic input covariance matrix at TX1. Note that, $\rank{\Hm_{21}}=N$ by construction, thus $\dimV{\ker{(\Hm_{21})}}=M-N$ and $\ker{(\Hm_{21})} \subseteq \Rc^{M \times (M-N)}$. Therefore, a linear precoder $\Zm_2 \in \mathbb{C}^{M \times (M-N)}$ such that $\spanV{(\Zm_2)} = \ker{(\Hm_{21})}$ can always be found. 

Now, we let $\uv_2\sim\Cc\Nc(0, d(p_{2,1},\dots,p_{2,(M-N)}))$ be an input vector of size $M-N$ such that $\xv_2=\Zm_2 \uv_2$, \linebreak $\Sm_2=\Zm d(p_{2,1},\dots,p_{2,(M-N)})\Zm\herm$ be the covariance matrix of $\uv_2$ and $P_{2}=\mathbb{E}[x_{i,j}x_{i,j}\herm]$ be the average transmit power per precoded symbol at TX2, $\forall j \in [1,M]$. Then, the maximum achievable spectral efficiency for the secondary link is the solution of the following maximization problem
\begin{eqnarray}\nonumber \label{eq:max_1}
    \max_{\Sm_2} & & \frac{1}{M} \log_{2}\left|\Id_N + \Sm_{(2,\xi)}^{-1/2}\Hm_{22}\Sm_2 \Hm_{22}\herm \Sm_{(2,\xi)}^{-1/2}\right|  \\ 
 \text{s.t.}  & &  \Hm_{21} \Zm_2 = \textbf{0}_{N \times (M-N)} \\ \nonumber & &  \trace(\Sm_2) \leq M P_{2}.
\end{eqnarray}
The presence of the constraint $\Hm_{21} \Zm_2 = \textbf{0}_{N \times (M-N)}$ restricts the subset of the possible solutions to the kernel of the interference channel.  Let $\Wm$ be a matrix whose columns form an orthonormal basis of $\spanV{(\Zm_2)}$. Such one $\Wm$ is semi-unitary by definition of orthonormal matrix and many strategies can be adopted to derive it, e.g. LQ factorization. Then, by defining $\Gammam \in \mathbb{C}^{(M-N) \times (M-N)}$ as a matrix with random entries, we can remove the constraint and write
\begin{equation} \label{eq:precoder_proof}
\Zm_2 = \Wm \Gammam.
\end{equation}
The columns of $\Zm_2$ are a generic linear combination of the columns of $\Wm$, thus $\Hm_{21} \Zm_2=\textbf{0}_{N \times (M-N)}$ will be satisfied by any optimal $\Zm_2^{*}=\Wm \Gammam^{*}$ by construction. Then we can write
\begin{equation} \label{eq:sigma}
\Sm_{2} =  \Wm \Gammam d(p_{2,1},\dots,p_{2,(M-N)}) \Gammam\herm \Wm\herm= \Wm \Sigmam_2 \Wm\herm
\end{equation}
with $\Sigmam_2 = \Gammam d(p_{2,1},\dots,p_{2,(M-N)}) \Gammam\herm$, and (\ref{eq:max_1}) becomes
\begin{IEEEeqnarray}{rCl}
\max_{\Sm_2} & & \quad \frac{1}{M} \log_{2}\left|\Id_N + \Sm_{(2,\xi)}^{-1/2}\Hm_{22}\Wm \Sigmam_2 \Wm\herm \Hm_{22}\herm \Sm_{(2,\xi)}^{-1/2}\right|  \nonumber \\
\text{s.t.}  & & \quad \trace(\Sigmam_2) \leq M P_{2}. \label{eq:max_11}
\end{IEEEeqnarray}
We further simplify (\ref{eq:max_11}), by letting $\Gm = \Sm_{(2,\xi)}^{-1/2}\Hm_{22}\Wm$. At this stage, we can take its SVD and write $\Gm~=~\Um_{\gx}\Lambdam_{\gx}^{\frac{1}{2}}\Vm_{\gx}^{\text{H}}$, with \linebreak $\Um_{\gx}\in\mathbb{C}^{N\times N}$,$\Vm_{\gx}\in\mathbb{C}^{(M-N) \times (M-N)}$ unitary matrices. Moreover, $\Lambdam_{\gx}=~[\Lambdam_{\gx}^{\lambda}, \Lambdam_{\gx}^{0}]^{\text{T}}$, where $\Lambdam_{\gx}^{0}=\textbf{0}_{(M-N) \times 2N-M}$ and $\Lambdam_{\gx}^{\lambda}=d(\sqrt{\lambda_{(\gx,1)}},\dots,\sqrt{\lambda_{(\gx,(M-N))}})$, with $\lambda_{(\gx,i)}$ eigenvalues of $\Gm\herm\Gm$. Therefore, we can write
\begin{eqnarray}\nonumber \label{eq:max_2}
    \max_{\Sigmam_2} & & \frac{1}{M} \log_{2}\left|\Id_N + \Um_{\gx}\Lambdam_{\gx}^{\frac{1}{2}}\Vm_{\gx}\herm \Sigmam_2 \Vm_{\gx}\Lambdam_{\gx}^{\frac{1}{2}} \Um_{\gx}\herm \right|  \\
 \text{s.t.}  & &   \trace(\Sigmam_2) \leq M P_{2}.
\end{eqnarray}
The upper bound for the determinant of a positive definite matrix is given by the product of the elements on its main diagonal, i.e., $|\Am| \leq \prod_i \Am_{[i,i]}$ (Hadamard inequality). Then, if we let $\Gammam = \Vm_\gx$, thus $\Sigmam_2 = \Vm_{\text{g}} d(p_{2,1},\dots,p_{2,(M-N)}) \Vm_{\gx}\herm$, the argument of the determinant in (\ref{eq:max_2}) is diagonalized and we can write
\begin{eqnarray}\label{eq:newproblem}
    \max_{p_{2,i}} & & \sum_{i=1}^{M-N}\log_{2}(1 +p_{2,i}\lambda_{(\gx,i)})  \\
    \text{s.t.} & & \sum_{i=1}^{M-N} p_{2,i} \leq M P_{2}. \nonumber
\end{eqnarray}
Now we can apply a classical water-filling algorithm to find 
\begin{equation}\label{eq:wf2}
p_{2,i} = \left[\mu -\frac{1}{\lambda_{(\gx,i)}}\right]^+,
\end{equation}
with $\mu$, the so-called ``water level'', determined such that \linebreak $\sum_i^{(M-N)} p_{2,i} \leq M P_2$. Now, it is clear that $\Gammam^{*}=\Vm_\gx$, and the solution to (\ref{eq:max_1}) is $\Sm_2 =~\Wm \Vm_{\gx} d(p_{2,1},\dots,p_{2,(M-N)}) \Vm_{\gx}\herm \Wm \herm$. By plugging $\Gammam^{*}$ in (\ref{eq:precoder_proof}), we obtain  
\begin{equation}
\Zm^* = \Wm \Vm_{\gx},
\end{equation}
where $\Wm$ is semi-unitary by construction and $\Vm_{\gx}$ is unitary by definition of SVD. The spectral efficiency maximizing precoder, under the considered constraints, is then semi-unitary\footnote{A unitary precoder $\Am \in \mathbb{C}^{N \times N}$ is a particular case of semi-unitary precoder, in fact $\rank{(\Am)}=min\{N,N\}=N$ and all its eigenvalues are equal to 1, thus Proposition \ref{prop:prod} can be applied.}. In particular, in \eqref{eq:precoder_proof}, $\Wm$ can be composed by any orthonormal set of columns spanning $\ker{(\Hm_{12})}$, whose appropriate linear combination to maximize the spectral efficiency will always be found by means of a suitable $\Vm_{\gx}$, thus $\Wm$ is optimal and this ends the proof.
\end{IEEEproof} 

\subsection{Proof of corollary \ref{corol:cia}} \label{sec:appendix_C}
\begin{IEEEproof} 
Consider \eqref{eq:y_p_bef_proc} and \eqref{eq:received_s2}. We are focusing on the link from the $k$th SBS to the $k$th SUE, thus, according to \eqref{eq:x_theta}, we can rewrite them as follows
\begin{equation} \label{eq:x_channel_thm}
\begin{array}{lcl}
\yv_{\px}^{(j)} = \Dm_j \Fm  \Bm \left(\Hm_{\text{pp}}^{(1,j)}\Am \Fm^{-1}\xv_{\px} + \nv_{\px}^{(j)}\right) + \Tm_{\sx\px}^{(k,j)} \Zm^{(k)}\uv_{\sx}^{(i)} \\
\yv_\sx^{(k)} = \Fm \Bm \left(\Hm_{\text{ss}}^{(k,k)}\Zm^{(k)} \uv_{\sx}^{(k)} + \Hm_{\px\sx}^{(1,k)} \xv_{\px} + \nv_{\sx}^{(k)}\right),
\end{array}
\end{equation}
where we omitted the co-tier interference component at the $k$th SUE, $\sum_{i=1, i \neq k}^{K} \Hm_{\text{ss}}^{(i,k)} \Zm^{(i)} \uv_{\sx}^{(i)}$, not known at the $k$th SBS, in compliance with the model described in Sec. \ref{sec:model}. Moreover, we isolated the cross-tier interference component coming from the $k$th SBS to the $j$th MUE for clarity. This comes without loss of generality, given that $\Tm_{\sx\px}^{(k,j)} \Zm^{(k)}=\textbf{0}$, $\forall j \in [1,M]$, $k \in [1,K]$, by construction. By looking at \eqref{eq:x_channel_thm}, we recognize the interference channel equations provided in the hypothesis of Proposition \ref{proposition:general}. Now we switch our focus on the cascaded precoder $\Zm^{(k)}= \Em^{(k)}\Thetam^{(k)}$. Consider CIA A and CIA B as described in Sec. \ref{sec:co_tier}. In the former, both $\Em^{(k)}$ and $\Thetam^{(k)}$ are semi-unitary by construction, hence so is $\Zm^{(k)}$ for Proposition \ref{prop:prod}, and Proposition \ref{proposition:general} can be applied. In the latter, $\Thetam^{(k)}$ is not semi-unitary but, by construction, selects the $\theta$ columns of $\Em^{(k)}$ associated to the $\theta$ indexes belonging to $\Nc^{\theta}$. Then, $\Zm^{(k)}$ is composed of orthonormal columns, thus it is semi-unitary by definition, and Proposition \ref{proposition:general} can be applied to conclude the proof.
\end{IEEEproof}

\bibliographystyle{unsrt}
\bibliography{distributed_CIA}

\vspace*{-2\baselineskip}

\begin{IEEEbiography}[{\includegraphics[width=1in,clip, keepaspectratio]{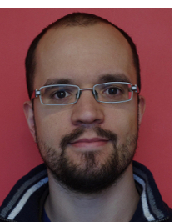}}]
{Marco Maso} received the bachelor’s degree in 2005 and the M.Sc. degree in Telecommunications Engineering in 2008, both from University of Padova, Italy, and is currently pursuing his Ph.D degree at University of Padova and Sup\'elec. He worked on projects dealing with practical implementations of OFDM packet synchronization  in 2005/06, and DVB-T2 system simulation in 2008/09. He is currently involved in the HENIAC project, studying new techniques for high speed coherent optical communications. His research interests include heterogeneous networks, wireless communications, cognitive radio and embedded devices.
\end{IEEEbiography}

\vspace*{-2\baselineskip}

\begin{IEEEbiography}[{\includegraphics[width=1in,clip, keepaspectratio]{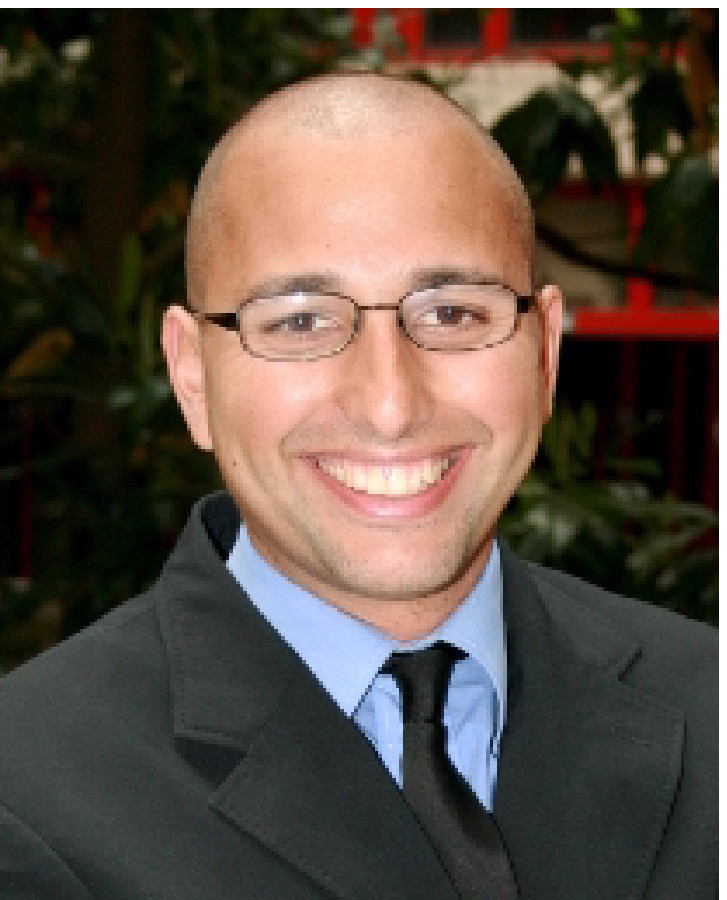}}]
{M\'erouane Debbah} entered the École Normale Sup\'erieure de Cachan (France) in 1996 where he received his M.Sc and Ph.D. degrees respectively. He worked for Motorola Labs (Saclay, France) from 1999-2002 and the Vienna Research Center for Telecommunications (Vienna, Austria) from  2002-2003. He then joined the Mobile Communications department of the Institut Eurecom (Sophia Antipolis, France) as an Assistant Professor. Since 2007, he is a Full Professor at Sup\'elec (Gif-sur-Yvette, France), holder of the  Alcatel-Lucent Chair on Flexible Radio. His research interests are in information theory, signal processing and wireless communications. He is an Associate Editor for IEEE Transactions on Signal Processing. M\'erouane Debbah is the recipient of the "Mario Boella" prize award in 2005, the 2007 General Symposium IEEE GLOBECOM best paper award, the Wi-Opt 2009 best paper award, the 2010 Newcom++ best paper award as well as the Valuetools 2007, Valuetools 2008 and CrownCom2009 best student paper awards. He is a WWRF fellow. In 2011, he received the IEEE/SEE Glavieux Prize Award.
\end{IEEEbiography}

\vspace*{-2\baselineskip}

\begin{IEEEbiography}[{\includegraphics[width=1in,clip, keepaspectratio]{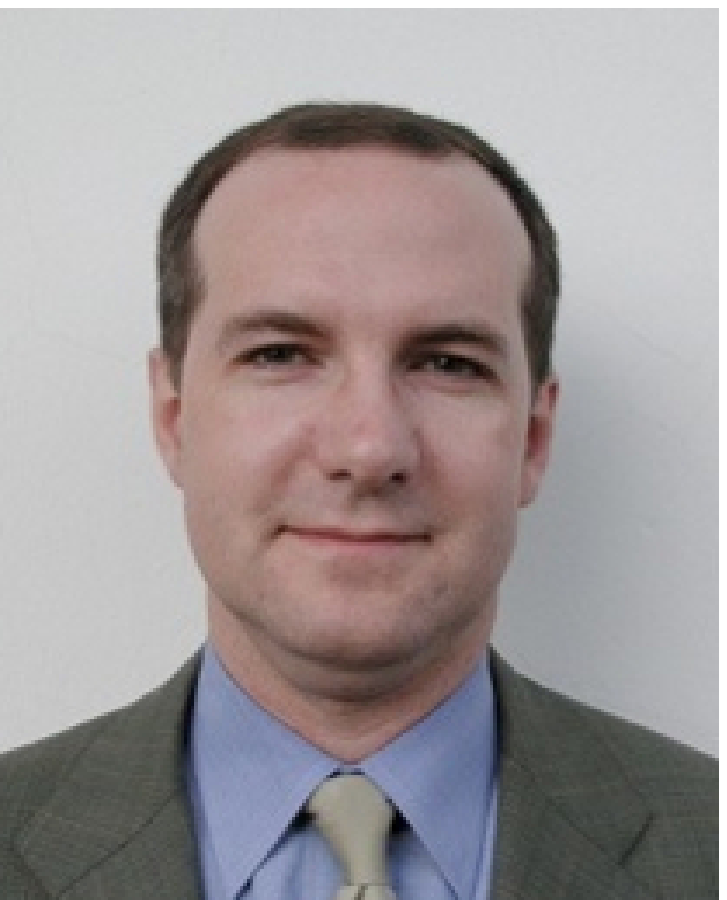}}]
{Lorenzo Vangelista} (SM’02) received the Laurea degree from University of Padova, Padova, Italy, in 1992, and the Ph.D. degree in Electrical and Telecommunication Engineering from University of Padova,in 1995. He subsequently joined the Transmission and Optical Technology Department of CSELT, Torino, Italy. From December 1996 to January 2002, he was with Telit Mobile Terminals, Sgonico (TS), Italy and then, up to May 2003, with Microcell A/S, Copenaghen, Denmark. Until July 2006 he has been with the worldwide organization of Infineon Technologies, as program manager. Since October 2006 he is an Associate Professor of Telecommunication within the Department of Information Engineering of Padova University, Italy. His research interests include signal theory, multi-carrier modulation techniques, cellular networks, wireless sensors and actuators networks and smart-grids.
\end{IEEEbiography}

\end{document}

%% file: macros.tex
\newfont{\bbb}{msbm10 scaled 500}

\newfont{\bb}{msbm10 scaled 1100}

\newcommand{\av}{{\bf a}}

\newcommand{\ev}{{\bf e}}

\newcommand{\gv}{{\bf g}}
\newcommand{\hv}{{\bf h}}

\newcommand{\nv}{{\bf n}}

\newcommand{\rv}{{\bf r}}
\newcommand{\sv}{{\bf s}}
\newcommand{\tv}{{\bf t}}
\newcommand{\uv}{{\bf u}}

\newcommand{\vv}{{\bf v}}
\newcommand{\xv}{{\bf x}}
\newcommand{\yv}{{\bf y}}

\newcommand{\zerov}{{\bf 0}}

\newcommand{\Am}{{\bf A}}
\newcommand{\Bm}{{\bf B}}

\newcommand{\Dm}{{\bf D}}
\newcommand{\Em}{{\bf E}}
\newcommand{\Fm}{{\bf F}}
\newcommand{\Gm}{{\bf G}}
\newcommand{\Hm}{{\bf H}}
\newcommand{\Id}{{\bf I}}

\newcommand{\Qm}{{\bf Q}}

\newcommand{\Sm}{{\bf S}}
\newcommand{\Tm}{{\bf T}}
\newcommand{\Um}{{\bf U}}
\newcommand{\Wm}{{\bf W}}
\newcommand{\Vm}{{\bf V}}

\newcommand{\Zm}{{\bf Z}}

\newcommand{\Ac}{{\cal A}}
\newcommand{\Bc}{{\cal B}}
\newcommand{\Cc}{{\cal C}}

\newcommand{\Nc}{{\cal N}}

\newcommand{\Rc}{{\cal R}}

\newcommand{\Tc}{{\cal T}}

\newcommand{\Wc}{{\cal W}}
\newcommand{\Vc}{{\cal V}}

\newcommand{\xiv}{\hbox{\boldmath$\xi$}}

\newcommand{\Gammam}{\hbox{\boldmath$\Gamma$}}
\newcommand{\Lambdam}{\hbox{\boldmath$\Lambda$}}

\newcommand{\Sigmam}{\hbox{\boldmath$\Sigma$}}

\newcommand{\Thetam}{\hbox{\boldmath$\Theta$}}

\newcommand{\trace}{{\hbox{tr}}}

\newcommand{\herm}{^{\text{H}}}

\newcommand{\ax}{{\text{a}}}
\newcommand{\bx}{{\text{b}}}

\newcommand{\gx}{{\text{g}}}

\newcommand{\px}{{\text{p}}}

\newcommand{\sx}{{\text{s}}}